\documentclass[12pt]{article}
\usepackage[utf8]{inputenc}
\usepackage{amssymb}
\usepackage{amsfonts}
\usepackage{amsmath}
\usepackage{mathrsfs}
\usepackage[margin=1in]{geometry}
\usepackage{lscape}
\usepackage[onehalfspacing]{setspace}
\usepackage{natbib}
\usepackage{pdfpages}
\usepackage{pdfpages}
\usepackage{xfrac}
\usepackage{float}
\usepackage{epstopdf}
\usepackage{graphicx}
\usepackage[position=top]{subfig}

\usepackage{bbm}
\usepackage{lscape}
\usepackage{subfig}
\usepackage{graphicx}
\usepackage{setspace}
\usepackage[flushleft]{threeparttable}
\usepackage{tabularx}

\usepackage{comment}

\usepackage[labelfont=bf]{caption}
\usepackage[sc]{mathpazo}
\usepackage[T1]{fontenc}
\usepackage{epstopdf}

\usepackage{standalone}
\usepackage{tikz}
\usepackage{algorithm}
\usepackage{algorithmic}
\usepackage{chngcntr} %

\usepackage{siunitx} %
\usepackage{tablefootnote} %
\usepackage{booktabs} %

\usepackage[colorlinks=true]{hyperref}
\hypersetup{
  colorlinks   = true, %
  urlcolor     = blue,  %
  linkcolor    = blue,  %
  citecolor    = blue   %
}

\makeatletter
\def\proceduretype{}
\newcommand\PROCEDURE[3][default]{%
  \def\proceduretype{#1}%
  \ALC@it
  \textbf{#1}\ \textsc{#2} {#3}%
  \begin{ALC@prc}%
}
\newcommand\ENDPROCEDURE{%
  \end{ALC@prc}%
  \ifthenelse{\boolean{ALC@noend}}{}{%
    \ALC@it\algorithmicend\ \textbf{\proceduretype}%
  }%
}
\newenvironment{ALC@prc}{\begin{ALC@g}}{\end{ALC@g}}
\makeatother

\begin{document}

\title{\textbf{Trade Wars with Trade Deficits}\thanks{We thank Bettina Brueggemann, Manolis Chatzikonstantinou, Jevan Cherniwchan, Matt Doyle, Alok Johri, Tim Kehoe, Adam Lavecchia, Oliver Loertscher, Zach Mahone, Colin Mang, Gajen Raveendranathan, and Kim Ruhl.  Pujolas thanks SSHRC for Insight Grants 435-2021-0006, 435-2024-0339, and for the Partnership to Study Productivity, Firms and Incomes. Pujolas: pujolasp@mcmaster.ca; Rossbach: Jack.Rossbach@georgetown.edu.}}
\author{\textbf{Pau S. Pujolas}  \\ \small{McMaster University} \and \textbf{Jack Rossbach} \\ \small{Georgetown University - Qatar}}

\date{
\begin{tabular}{ccc}
&&\\
&December 2024&
\end{tabular}}
\maketitle
\thispagestyle{empty}

\begin{abstract}
\noindent
Trade imbalances significantly alter the welfare implications of tariffs. Using an illustrative model, we show that trade deficits enhance a country’s ability to alter its terms of trade, and thereby benefit from tariffs. Greater trade deficits imply higher optimal, or welfare maximizing, tariffs. We compute optimal unilateral and Nash equilibrium tariffs between the United States and China --- the countries with the largest bilateral trade imbalance --- using a multi-region, multi-sector applied general equilibrium model with service sectors and input-output linkages, a computationally complex task.  Free trade benefits both countries compared to a trade war.  Relative to existing tariff rates, however, the United States gains from a trade war with China --- a result that hinges on their bilateral trade imbalance.

\textbf{JEL\ Codes}: F11, F13, F14, F17
\end{abstract}

\newpage
 \pagenumbering{arabic}

\section{Introduction}
\label{sec:introduction}
Debate over the importance of international trade imbalances remains a perennial topic in political discourse and the media, with large trade deficits frequently cited as justification for protectionist policy measures.\footnote{\cite*{delpeuchTradeImbalancesFiscal2024} observe this connection is more than just rhetoric, finding that bilateral imbalances are associated with an increase in protectionist measures.}  This connection was particularly evident in the recent U.S.-China trade war, in which the trade imbalance between the United States and China was cited as one of the primary motivations for the initial round of tariff escalations.\footnote{The text reads \textit{``China has consistently taken advantage of the American economy [...] The United States has run a trade in goods deficit with China for years, including a \$375 billion deficit in 2017 alone."} \cite{DJT2018}.}  Despite the centrality of trade imbalances in political rhetoric and public perception of the trade war, the issue has received comparatively little attention in academic studies.\footnote{For example, two of the most cited review studies  evaluating the economic impact of the U.S.-China trade war contain no references to or discussion of trade imbalances or deficits (\citeauthor*{amitiImpact2018Tariffs2019}, \citeyear{amitiImpact2018Tariffs2019}; \citeauthor*{fajgelbaumEconomicImpactsUS2022}, \citeyear{fajgelbaumEconomicImpactsUS2022}).}

In this paper we study how trade imbalances impact the welfare effects of trade policy, which we examine both theoretically and quantitatively. To illustrate the interaction between trade policy and trade imbalances, we revisit the classic result on optimal tariff policy by \cite{johnsonOptimumTariffsRetaliation1953} and extend it to a context with trade deficits. We demonstrate that a trade deficit makes foreign demand more inelastic, enabling the home country to better manipulate the terms of trade. This interaction alters the welfare impact of tariffs, resulting in higher optimal, or welfare-maximizing, tariff rates.  Through numerical simulations, we show that these findings hold both when bilateral trade imbalances generate aggregate trade imbalances and when there are only bilateral --- but not aggregate --- trade imbalances.

Although this result might initially seem surprising, given that trade deficits are sometimes framed as an economic weakness, the underlying intuition is straightforward and matches what one might expect in a partial equilibrium environment.  It has long been understood that unilateral tariffs can be welfare improving \citep{costinotChapterTradeTheory2014}.  This is because the cost of the tariff is shared between the importer and exporter, whereas the benefit of the tariff --- the tariff revenue --- accrues solely to the importer.  Since a country with a trade deficit imports more than it exports, ceteris paribus, it benefits from higher tariff revenue and faces less offsetting retaliation from its trading partner.  As trade imbalances increase, this asymmetry becomes more pronounced, further favoring the country with a trade deficit.

We explore the quantitative importance of the interaction between trade imbalances and optimal tariff policy within a multi-region, multi-sector applied general equilibrium model featuring input-output linkages and endogenous bilateral trade deficits.\footnote{\citet*{Allen2019} develop general conditions for computing counterfactual predictions for the class of general equilibrium gravity models, which similarly allow for trade deficits.  They show that counterfactual predictions for these models rely only on demand and supply elasticities.  They estimate the impact of a flat 10 percent increases in bilateral iceberg trade frictions between the U.S. and China and find small decreases in real output prices and larger decreases in real expenditures.  They note, however, that models that include tariffs do not generally fall within the class of universal gravity models, since tariffs serve as an additional source of domestic income.  Unlike tariffs, iceberg trade costs do not generate income and therefore have markedly different welfare implications.}  As \citet*{kehoeQuantitativeTradeModels2017} discuss, these models remain the principal framework used by policymakers to assess the economic impact of changes in international trade policy.\footnote{Our baseline analysis follows the tradition of Applied or Computable General Equilibrium models built using an Armington structure. In Appendix \ref{appendix:CP} we repeat our analysis using the multi-sector \citet{eatonTechnologyGeographyTrade2002} Ricardian framework from \citet{caliendoEstimatesTradeWelfare2015} and achieve similar results.}  Our motivating application is a trade war between the United States and China, inspired by the eruption in trade tensions between the two countries that started when, in January 2018, the United States announced tariff hikes on imports from China.  This conflict is notable for our framework because China is the largest exporter to the United States, the United States is the largest trading partner of China, and the trade imbalance between them is the largest across all bilateral country pairs.  

We calibrate our model to match observed trade flows and tariffs in the data and compute optimal unilateral and Nash equilibrium tariffs between the United States and China.   We find that the United States experiences welfare gains in both the unilateral and Nash equilibrium cases compared to the pre-trade war baseline.  This result is driven both by the large bilateral deficit that the United States has with China, and by China's relatively high baseline tariffs on imports from the United States --- both of which reduce the potential loss from retaliation.  In particular, when we eliminate bilateral trade imbalances between the United States and China, both countries lose from a trade war relative to existing tariffs.  Compared to a baseline with free trade, the United States experiences small welfare losses. This conclusion holds only when both services and input-output linkages are included in our model.  This result highlights the importance of conducting trade policy analysis in a fully-specified framework, as emphasized by \citet{caliendoEstimatesTradeWelfare2015}.  We further evaluate the welfare impact of trade wars with other trading partners, and find that the United States would benefit from starting trade wars --- by which we mean moving to the Nash equilibrium tariff rates --- with a number of its trading partners.  These results might help rationalize the increasingly antagonistic stance of countries with deficits in international trade disputes  \citep*{delpeuchTradeImbalancesFiscal2024}.   

The reason relatively little attention has been given to the interaction between trade imbalances and international trade policy is likely because trade imbalances are understood first and foremost as a dynamic consideration that arises from movements in domestic fiscal policy, rather than a phenomenon to be explained by international trade policy \citep{Krugman1986}.\footnote{\cite{DixCarneiroEtAl} and \cite{Carneiro23} are notable exceptions that focus on the connection between trade imbalances and labor markets.} Supporting this view,  \citet{furceriMacroeconomicConsequencesTariffs2018} use a reduced form approach to study the impact of the U.S.-China trade war on a variety of outcomes, including the trade balance, and find that the trade war did little to impact trade imbalances.  This result is consistent with our modeling approach, and does not imply that trade imbalances cannot alter the effects of trade policy.

In dynamic models, aggregate trade imbalances arise endogenously as a means for countries to share risk and smooth consumption across periods.\footnote{See \cite*{BKK} and \cite*{BKK2} for early quantitative models to feature endogenous trade imbalances, and \cite{PerriFogli}, \citet*{kehoeGlobalImbalancesStructural2018}, or \cite{Joe19RED}  for more recent work.}  These frameworks, however, typically do not distinguish between aggregate and bilateral trade deficits, and imbalances are assumed to balance out over time, which may limit the room for trade policy to interact with transient trade imbalances. For instance, \cite{BESHKAR202065} study optimal inter-temporal trade policy in a two-country dynamic Ricardian model, in which trade imbalances arise endogenously and function to smooth consumption over time.  They show that optimal trade policy is counter-cyclical, with optimal tariffs and capital controls in each period uniquely determined by relative productivities in that period. In their framework, trade imbalances are not generally persistent, do not impact the share of production consumed at home and abroad in a given period, and have no direct impact on optimal trade policy.  In contrast, in a static framework, trade imbalances necessarily correspond to imbalances in the shares of production consumed across countries.  A dynamic model that features persistence in trade imbalances, similar to the persistence we observe in the data (see \citeauthor{lubikShouldWeWorry2017}, \citeyear{lubikShouldWeWorry2017}; or \citeauthor*{kehoeGlobalImbalancesStructural2018}, \citeyear{kehoeGlobalImbalancesStructural2018}), might lead to trade imbalances impacting consumption shares and thereby create space to derive results similar to those in our static framework.

To our knowledge, ours is the first paper to provide optimal tariffs in a fully-specified applied general equilibrium framework with many sectors and input-output linkages across sectors.  This is likely due to the computational complexity of calculating Nash equilibria in higher dimension models.  For example, \cite{heArmingtonAssumptionSize2017a} compute optimal tariffs between two countries in a two-sector, three-country framework where the third country is a ROW aggregate.  In contrast, our model has 18 regions and 22 tradable goods-producing sectors.  In such a framework, it is unfeasible to compute optimal tariffs through an exhaustive search, since the strategy space grows polynomially with the number of sectors.  For example, if we discretize the solution space into 10 possible tariff rates per sector, a three sector model has 1000 possible strategies for each country, whereas a 22 sector model has $10^{22}$ or 10 Sextillion such strategies.\footnote{In our application, we discretize the solution space into 40,000 possible tariff rates for each sector.} We employ a genetic algorithm, which is a global optimization heuristic, to find optimal tariff rates for each country in response to the other country's tariffs.  We then iterate these best-response tariff rates until we converge to a Nash equilibrium, where neither country wants to deviate.  While optimal tariff rates vary across sectors and specifications, in our application to the U.S.--China trade war, we find the average optimal tariff rate is somewhat stable across specifications and in the range of 7--16 percent.  This range is generally consistent with optimal tariff rates found elsewhere in the literature, for example by \citet{caliendoSecondbestArgumentLow2023}.

After evaluating the impact of a trade war in which tariffs are set to maximize welfare, we use our model to measure the welfare implications of the 2018 tariff escalations between the United States and China observed in the data.  We find that both the United States and China experienced welfare losses from their recent trade war.\footnote{In Appendix \ref{subsec:uniform_tariffs} we quantitatively evaluate the United States setting a uniform tariff rate on its trading partners.  We find that the United States gains from unilaterally imposing such tariffs; however, if countries respond to U.S. tariffs in equal measure, all countries  experience welfare losses from the resulting global trade war.}  These losses are consistent with a large number of papers that have calculated the impact of the 2018 tariffs, such as \cite*{amitiImpact2018Tariffs2019}, \cite*{FlaaenHortacsuTintenot}, or \cite{carrollHur23}.\footnote{We refer readers to \cite{fajgelbaumEconomicImpactsUS2022} for a broad overview of the trade conflict and a review of the numerous studies of its economic impact, the majority of which find small welfare losses for both the United States and China.}  We find that U.S. tariff increases during the trade war were negatively correlated with the increases consistent with a move towards Nash equilibrium tariff rates.  This provides suggestive evidence in favor of arguments by \citet{Fetzer2021} and \citet{KimMargalit2021} who claim that the tariffs were politically, rather than economically, motivated.

\section{Illustrative Two-Country Model}
\label{sec:toy_model}
In this section we develop a simple, two-country \citet{Armington1969} model with CES preferences to highlight the interaction between trade imbalances and optimal tariff policy.  Our environment is an extension of the canonical framework of \cite{johnsonOptimumTariffsRetaliation1953}, in which optimal tariffs are shown to be the inverse of the other country's domestic absorption and trade elasticity --- a result that we recover when there are no imbalances.   %

We consider an environment where there are two countries and two goods.  Each country values both goods but is only endowed with a positive amount of one of them.  Specifically, Country 1 is endowed with $E_1$ units of good 1, and Country 2 is endowed with $E_2$ units of good 2.

Country 1's preferences are given by $U^1(c_1^1,c_2^1)$, where $c_1^1$ is Country 1's consumption of good 1, and $c_2^1$ is consumption of good 2.  All of good 2 is imported, while consumption of good 1 is domestic production minus exports to Country 2, $c_1^1=E_1-c_1^2$. There is an exogenous trade deficit $d$, which implies that the trade balance in this economy is given by $pc_2^1=c_1^2+d$, where $p$ is the price of good 2 in terms of good 1.\footnote{It is possible for transfers to harm the beneficiary country in general equilibrium models, an effect known as the transfer paradox (see \citeauthor{LEO36}, \citeyear{LEO36}; \citeauthor{samuelson52}, \citeyear{samuelson52}; or \citeauthor{gale74}, \citeyear{gale74}; among many others). While $d$ in our model appears similar to the transfer in those results, this effect is not what drives our results.  In later sections of the paper, we show that our results on optimal tariff setting hold even when we eliminate aggregate trade imbalances, such that only bilateral imbalances remain --- and therefore no country receives a ``transfer."}  The government in Country 1 can impose an import tariff $\tau_1$, which will affect the relative price of the foreign good. 

The price-taking representative consumer in Country 1 faces a tariff on imports, $\tau_1$, and solves the following maximization problem,
\begin{equation*}
\begin{aligned}
&\max U^1(c_1^1,c_2^1)\\
\text{subject to:  }& c_1^1+p(1+\tau_1)c_2^1=E_1+d+T_1,\\
\end{aligned}
\end{equation*}
where $T_1=p\tau_{1}c_2^1$.  Let $U^1_i$ represent the derivative of the utility function with respect to good $i$.  This implies that the allocation satisfies  
\begin{equation}
\label{eq1toy}
U^1_2=p(1+\tau_1)U^1_1.
\end{equation}

In the same environment, we consider a benevolent planner seeking to maximize the welfare of consumers in Country 1. The planner's allocations affect relative prices, $p$, which it takes into consideration. Therefore, the benevolent planner solves the problem 
\begin{equation*}
\begin{aligned}
&\max U^1(c_1^1,c_2^1)\\
\text{subject to:  }& c_1^1=E_1-pc_2^1+d\\
\end{aligned}
\end{equation*}
which implies that
\begin{equation}
\label{eq2toy}
U^1_2=pU^1_1\left(1+\frac{d\log p}{d\log c_2^1}\right).
\end{equation}
A government setting the tariff, $\tau_1$, to maximize the welfare of consumers in Country 1, would accomplish this by finding the tariff such that equations (\ref{eq1toy}) and (\ref{eq2toy}) are equal.  We refer to this welfare-maximizing tariff as the optimal tariff.  This is the classical result in \cite{johnsonOptimumTariffsRetaliation1953}, in which the optimal tariff is equal to the inverse of its import elasticty,
\begin{equation}
\label{tariff1}
\tau_1=\frac{1}{\left(\frac{d\log c_2^1}{d\log p}\right)}={\frac{d\log p}{d\log c_2^1}}.
\end{equation}

The equivalent problem for Country 2 implies a household choice satisfying $pU^2_1=(1+\tau_2)U_2^2$ (note that in Country 2, good 1 costs $(1+\tau_2)$ and good 2 costs $p$, and their budget constraint is $(1+\tau_2)c_1^2+pc_2^2=pE_2-d+T_2$). Similarly, the benevolent planner's problem for Country 2 is  
$$
\max U^2(c_1^2,c_2^2) = \max U^2\left(c_1^2,E_2-\left(\frac{c_1^2+d}{p}\right)\right)
$$
implying that $pU^2_1=\left(1-\frac{c_1^2+d}{c_1^2}\frac{d\log p}{d\log c_1^2}\right)U_2^2$. Country 2's optimal tariff is then
\begin{equation}
\label{tariff2}
\tau_2=-\frac{c_1^2+d}{c_1^2}\frac{d\log p}{d\log c_1^2}=-\frac{pc_2^1}{c_1^2}\frac{d\log p}{d\log c_1^2}.
\end{equation}

Equation (\ref{tariff2}) also contains the inverse of its import elasticity (note that Country $2$ is importing good $1$). Equation (\ref{tariff2}) differs from equation (\ref{tariff1}) in two key ways. First, it has a negative sign: while $p$ is the price of Country 1's imports, it is not the price of Country 2's imports, but rather its exports. Second, the trade imbalance (trade surplus for Country 2) appears in equation (\ref{tariff2}) but not in equation (\ref{tariff1}). Since Country 2 is endowed only in units of good 2 while the deficit is in terms of good 1, Country 2 can affect the value of its trade surplus by manipulating the terms of trade.\footnote{To consider the case where the trade imbalance is instead valued in terms of the good of the country with the trade surplus, simply consider $d$ to be a negative number.} %

To be able to derive closed-form solutions, we continue by specifying constant elasticity of substitution (CES) preferences. Namely, Country $i$'s preferences are given by
$$
U^i(c_1^i,c_2^i) = \left(c_1^i\right)^{\frac{\sigma_i-1}{\sigma_i}}+\left(c_2^i\right)^{\frac{\sigma_i-1}{\sigma_i}}.
$$

We can solve for Country 2's relative demand of good 1 as a function of good 2,
\begin{equation}
\label{eqA}
c_1^2=c_2^2\left(\frac{p}{1+\tau_2}\right)^{\sigma_2},
\end{equation}
and for the demand of good 2, 
\begin{equation}
\label{eqB}
c_2^2=\frac{pE_2-d+T_2}{p\left(1+\left(\frac{p}{1+\tau_2}\right)^{\sigma_2-1}\right)}.
\end{equation}
We then proceed with the feasibility condition for good 2, $E_2=c_2^2+c_2^1$ to write Country 1's demand for good 2 as
\begin{equation}
\label{eq3toy}
c_2^1 = \frac{pE_2\left(\frac{p}{1+\tau_2}\right)^{\sigma_2-1}+d-T_2}{p\left(1+\left(\frac{p}{1+\tau_2}\right)^{\sigma_2-1}\right)}.
\end{equation}

We take the derivative of equation (\ref{eq3toy}), express it in logarithms, and combine it with equations (\ref{eqA}) and (\ref{eqB}) to find that
\begin{equation}
\label{eq4toy}
\frac{d\log c_2^1}{d\log p} = \left(\frac{E_2}{c_2^1}-1\right)
(\sigma_2-1)\lambda_2\left(\frac{p}{1+\tau_2}\right)^{\sigma_2-1}-\lambda_2\frac{d-T_2}{pc^1_2},
\end{equation}
where $\lambda_2$ is the domestic absorption of Country 2,
$$\lambda_2=\frac{pc_2^2}{pE_2-d+T_2}=\frac{1}{1+\left(\frac{p}{1+\tau_2}\right)^{\sigma_2-1}}.$$
Using this definition and the trade balance, we can write the expression for $\tau_1$, given by equation (\ref{tariff1}), as
\begin{equation}
\label{tariff1final}
\tau_1=\frac{1}{\lambda_2\left((\sigma_2(1+\tau_2)-1)\frac{c^2_1}{c^2_1+d}-\frac{d}{pc^1_2}\right)}.
\end{equation}

Equation \eqref{tariff1final} shows that the optimal tariff rate in Country 1, $\tau_1$, is inversely related to Country 2's domestic absorption, $\lambda_2$ and its demand elasticity for imports from Country 1, $\sigma_2$. In fact, if we set $\tau_2=d=0$, then equation \eqref{tariff1final} becomes the textbook case $\tau_1=(\lambda_2(\sigma_2-1))^{-1}$.  We further find that optimal tariffs set by Country 1 react to the tariffs raised by Country 2, $\tau_2$. In particular, a higher $\tau_2$ lowers $\tau_1$. %

Our key result in equation \eqref{tariff1final} is that a trade deficit, $d>0$, implies higher optimal tariffs for the country with the deficit, since: $c^2_1/(c^2_1+d)<1$. This term tampers the impact of the trade elasticity and Country 2's tariffs. Whereas a high trade elasticity and high tariffs from Country 2 lower Country 1's optimal tariff, a trade deficit reduces this effect, implying a higher optimal tariff rate.  Moreover, a trade deficit increases the demand for $c_2^1$, as shown in equation (\ref{eq3toy}). Country 1 can affect this channel by lowering $p$, and the greater Country 1's deficit is relative to total imports, $d/pc^1_2$, the more important this channel becomes.

\section{Global Trade Imbalances}

\label{sec:imbalances}
Prior to examining the implications of trade imbalances for trade policy, we document patterns in the evolution of global trade imbalances over time.  We report bilateral imbalances in each period as the absolute value of net exports divided by gross trade flows, where all objects are measured for a given bilateral relationship.  Specifically we compute

\begin{equation}
\text{Imbalance}_{ijt} = \frac{\left|\text{Flow}_{ijt} - \text{Flow}_{jit}\right|}{\text{Flow}_{ijt} + \text{Flow}_{jit}},
    \label{eq:imbalance}
\end{equation}
where $\text{Flow}_{ijt}$ is the trade flow from country $j$ to country $i$ in year $t$.  If trade is perfectly balanced for each bilateral relationship, net exports are zero, which implies that $\text{Imbalance}_{ijt}$ in equation (\ref{eq:imbalance}) is also zero.  At the other extreme, if a trade flows occurs only in one direction (a country either only imports or only exports), then $\text{Imbalance}_{ijt}$ equals 1.  Hence, our measure for bilateral imbalances is between 0 and 1, where 0 indicates perfect balance and 1 indicates complete imbalance.\footnote{Our measure of $\text{Imbalance}_{ijt}$ is reminiscent of the Grubel-Lloyd index of international trade (after \citeauthor{GrubelLloyd}, \citeyear{GrubelLloyd}); instead of looking at the composition of intra-industry trade for a country, we look at the composition of trade balances worldwide.}

\begin{figure}[htp]
\caption{Bilateral Imbalances as Percentage of Gross Trade Flows \label{Fig--Global_Imbalances} } %
	\begin{minipage}[b]{\linewidth}
		\centering
		 	\includegraphics[scale=0.9]{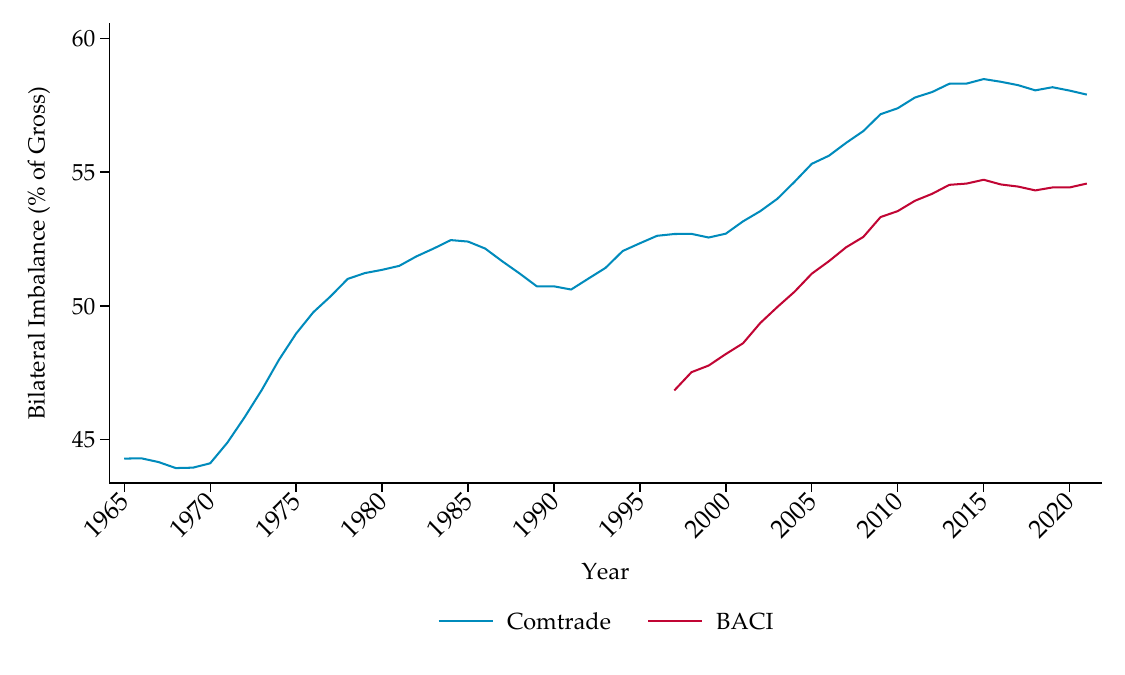}
	\end{minipage}

\end{figure}

Figure \ref{Fig--Global_Imbalances} shows the evolution of bilateral imbalances, reported as percentages, from 1965--2020, according to two databases: the UN Comtrade database, and the BACI database described by \citet{Gaulier2010};\footnote{For the Comtrade data, we use the trade flows as reported by the importer, except where not available.  The BACI database implements various techniques meant to reconcile discrepancies in reported trade flows in the underlying Comtrade data.} where the series are accessed through the CEPII Gravity Database \citep{Conte2022}.  For each series, we compute the weighted average of equation \eqref{eq:imbalance} across all bilateral pairs in a given year, where we weight by destination GDP, and then compute a 5 year moving average.  Figure \ref{Fig--Global_Imbalances}  shows that bilateral imbalances grew throughout the period.  Although the datasets give us a limited window for examining the implications of recent increases in global trade tensions, the plateau or reversal in global imbalances over the past decade is minimal in the context of broader trends.\footnote{The 5-year moving averages remain largely unaffected if we exclude 2020 to avoid the fall in trade that coincided with the global COVID-19 pandemic.}

Our results are consistent with those of \citet{Eugster2020} and \citet{Cunat2023}, who find that bilateral imbalances are large and persistent.  These papers examine the extent to which trade imbalances respond to or can be explained by trade policy --- a topic largely orthogonal to our focus on how trade imbalances alter the welfare implications of trade policy.  For example, \citet{Cunat2023} examine the unobserved iceberg trade costs necessary to rationalize bilateral trade imbalances and evaluate the welfare implications of removing these trade barriers.\footnote{In Appendix \ref{subsec:iceberg} we consider the implications of iceberg trade costs and their removal in our context of optimal tariff policy.}  \citet{Eugster2020} argue that changes in bilateral imbalances are best explained by macro factors and are not responsive to changes in trade policy.  They focus on explaining changes over 5 year periods between 1995--2015, which Figure \ref{Fig--Global_Imbalances} shows is a period with steadily increasing trade imbalances and is also a period over which tariffs were generally decreasing. %

\begin{figure}[htp]
\caption{Mean Aggregate Imbalance as Percentage of Gross Domestic Product \label{Fig--Global_Imbalances_GDP_Agg} } %
	\begin{minipage}[b]{\linewidth}
		\centering
		 	\includegraphics[scale=0.9]{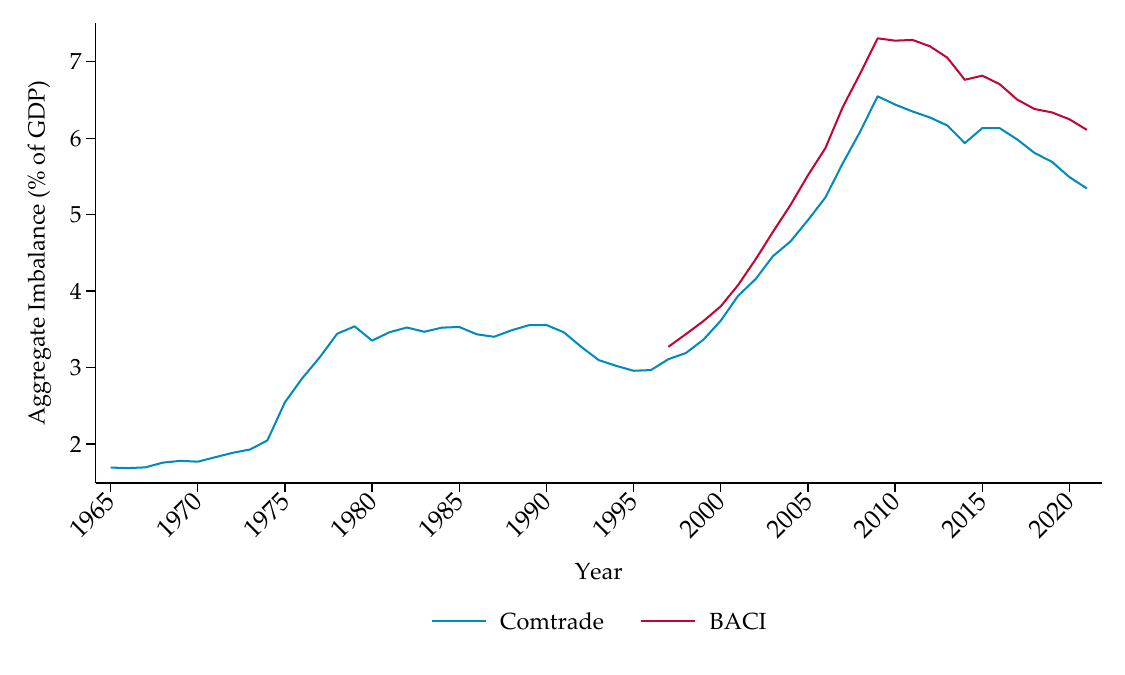}
	\end{minipage}

\end{figure}

We further report changes in aggregate trade imbalances, which we compute as
\begin{equation}
  \text{Agg.Imbalance}_{it} = \frac{\left|\sum_{j}{\left(\text{Flow}_{ijt} - \text{Flow}_{jit}\right)}\right|}{\text{GDP}_{it}}.
        \label{eq:agg_imbalance}
    \end{equation}
Figure \ref{Fig--Global_Imbalances_GDP_Agg} shows the evolution of aggregate trade imbalances according to the same two datasets we used to report bilateral imbalances.  We again weight imbalances by destination GDP when computing the average value in a given year, before plotting the 5 year moving average.  Note that equation \eqref{eq:agg_imbalance} can be decomposed into two effects: the first is the aggregate trade imbalance as a percentage of total trade flows, and the second is total trade flows as a percentage of GDP.  Therefore, this figure captures that net importers (exporters) are importing (exporting) more relative to their total trade flows, and that trade is increasing as a fraction of GDP.  The plot looks similar if one were to plot the aggregate imbalance as a percentage of total trade, rather than GDP, except with a smaller magnitude increase from 1995 onward.

Although there are small decreases in both bilateral and aggregate trade imbalances coinciding with recent increases in protectionism, their magnitudes are small and fall short of a full reversal.  Taken together, these results establish that bilateral and aggregate imbalances are large, and generally growing.  These imbalances, which have been associated with increases in protectionism  \citep{delpeuchTradeImbalancesFiscal2024}, motivate our analysis of the interaction between trade imbalances and the welfare impact of tariffs.

\section{Quantitative Model}
\label{sec:model}
In this section, we develop a multi-region, multi-sector applied general equilibrium trade model with input-output linkages, following \citet*{kehoeQuantitativeTradeModels2017}. Each country consists of households that consume final goods in each sector.  These goods are produced by perfectly competitive bundlers that combine domestic and foreign differentiated goods in that sector according to a CES specification. Tradable differentiated goods are produced using labor and final goods from each sector. Preferences and production technologies vary across countries and sectors.  Imported foreign differentiated goods may incur additional costs due to tariffs or iceberg trade costs, which vary by origin and sector.  Tariffs generate tariff revenue that is rebated back to households as a lump sum.  We describe the competitive equilibrium for this economy, which will later be used to analyze the problem of a government setting tariffs to maximize the welfare of domestic consumers.

\subsubsection*{Households}
\label{HHbaseline}
The representative household in country  $i$  solves

\begin{equation}
\label{Utility}
\max \sum_{s=1}^{S}a_{i}^{s}\log c_{i}^{s}
\end{equation}
subject to the budget constraint 
\begin{equation}
\label{BC}
\sum_{s=1}^S p_{i}^{s}c_{i}^{s}= {I}_i,
\end{equation}
where
\begin{equation}
\label{BC2}
 {I}_i \equiv w_{i}L_{i}+T_{i}+D_{i},
\end{equation}
and  \( s \)  is the sector, \(w_{i}L_{i}\) is labor income,  \( T_{i} \)  is tariff revenue, and  \( D_{i} \) is the aggregate trade balance, where \( D_{i} > 0\) indicates an aggregate trade deficit.  This problem has the solution
 \[ c_{i}^{s}=a_{i}^{s}\frac{{I}_i}{p_{i}^{s}}. \]

\subsubsection*{Final Goods Bundler}

The perfectly competitive final goods bundler in country  \( i \) for sector $s$ combines differentiated goods within the same sector from each country according to the Armington \citep{Armington1969} specification,
\begin{equation}
\label{Ydef}
y_{i}^{s}= \left(  \sum _{j=1}^{J} \gamma _{ij}^{s} \left( y_{ij}^{s} \right) ^{\frac{ \sigma _{s}-1}{ \sigma _{s}}} \right) ^{\frac{ \sigma _{s}}{ \sigma _{s}-1}},    
\end{equation} 
where  \(  \gamma _{ij}^{s} \)  is the sector weight for consumption in country  \( i \)  originating from country  \( j \)  in sector  \( s \), and  \(  \sigma _{s} \)  is the elasticity for sector  \( s \).

The bundler for sector  \( s \)  solves
 \[ \min \sum _{j=1}^{J}p_{ij}^{s}y_{ij}^{s} \] 
subject to the production function for \( y_{i}^{s} \) in equation (\ref{Ydef}).  This problem has the solution,
\begin{equation}
y_{ij}^{s}=(\gamma _{ij}^{s})^{\sigma_s}\left(\frac{p_{i}^{s}}{p_{ij}^{s}}\right)^{\sigma_s}y_{i}^{s},
\label{eq:yij_def}
\end{equation}
where  \( p_{i}^{s} \)  is the price index for sector  \( s  \) in country  \( i \),
\begin{equation}
    p_{i}^{s} \equiv \left(\sum_{s=1}^{s}  (\gamma _{ij}^{s})^{\sigma_s} (p_{ij}^{s})^{1-\sigma_s}\right) ^ \frac{1}{1-\sigma_s}.
    \label{eq:pi_def}
    \end{equation}
Note that the sector output,  \( y_{i}^{s} \), can be used as either final consumption,  \( c_{i}^{s} \), or as intermediate inputs,  \( x_{i}^{s} \).

\subsubsection*{Tradable Differentiated Goods Producers}

Tradable differentiated goods are produced according to the production function,

\begin{equation}
   y_{ij}^{s}=\frac{1}{t_{ij}^{s}} \min \left\{ z_{j}^{s}l_{ij}^{s},\frac{x_{ij}^{s1}}{ \alpha _{j}^{s1}}, \ldots \frac{x_{ij}^{sk}}{ \alpha _{j}^{sk}}, \ldots ,\frac{x_{ij}^{sS}}{ \alpha _{j}^{sS}} \right\},  
   \label{eq:prd_fcn}
  \end{equation}
where ${t_{ij}^{s}}$ is an iceberg trade cost, for example, due to non-tariff trade barriers or freight costs, and $z_{j}^{s}$ is productivity.  This means labor is given by

\begin{equation}
l_{ij}^s=t_{ij}^s\frac{y_{ij}^s}{z_{j}^s}.
\label{eq:lij_sol}
\end{equation}
The Leontief production function means that in equilibrium we will have
 \[ z_{j}^{s}l_{ij}^{s}=\frac{x_{ij}^{s1}}{ \alpha _{j}^{s1}}= \ldots =\frac{x_{ij}^{sk}}{ \alpha _{j}^{sk}}= \ldots =\frac{x_{ij}^{sS}}{ \alpha _{j}^{sS}}, \] 
which implies
 \[ l_{ij}^{s}=\frac{x_{ij}^{sk}}{ \alpha _{j}^{sk}}\frac{1}{z_{j}^{s}}. \]

 Importing foreign differentiated goods incurs ad valorem tariffs equal to  \(  \tau_{ij}^{s} - 1 \).  Prices, $p_{ij}^{s}$, are inclusive of tariffs, therefore the firm's profit maximization problem is 
 \[ \max\frac{p_{ij}^{s}y_{ij}^{s}}{ \tau_{ij}^{s}}-w_{j}l_{ij}^{s}- \sum _{k=1}^{S}p_{j}^{k}x_{ij}^{sk}. \] 
Fixed proportions in the production function allows us to write the problem only in terms of  \( l_{ij}^{s} \),
 \[ \max\frac{p_{ij}^{s}z_{j}^{s}l_{ij}^{s}}{ \tau_{ij}^{s}t_{ij}^{s}}-w_{j}l_{ij}^{s}- \sum _{k=1}^{S}p_{j}^{k} \alpha _{j}^{sk}z_{j}^{s}l_{ij}^{s}, \]
the solution of which implies that
 \[ z_{j}^{s}l_{ij}^{s} \left( \frac{p_{ij}^{s}}{ \tau_{ij}^{s}t_{ij}^{s}}- \sum _{k=1}^{S}p_{j}^{k} \alpha _{j}^{sk} \right) -w_{j}=0. \]
Rearranging shows that prices can be written in terms of trade costs, wages, productivity, and a weighted index of intermediate input prices,
\begin{equation}
  p_{ij}^{s}= \tau_{ij}^{s}t_{ij}^{s} \left( \frac{w_{j}}{z_{j}^{s}}+ \sum _{k=1}^{S}p_{j}^{k} \alpha _{j}^{sk} \right).  
\label{eq:pij_sol}
\end{equation}

\subsubsection*{Government Budget Constraint}
For country  \( i \) , the total tariff revenue rebated is equal to the tariff rates applied to import volumes, summed across all trading partners,

\begin{equation}
T_i=\sum_{s=1}^S\sum_{j=1}^J(\tau_{ij}^s-1)\frac{p_{ij}^sy_{ij}^s}{\tau_{ij}^s}.
\label{eq:tariffrevenue}
\end{equation}

\subsubsection*{Aggregate Trade Balance}

For country  \( i \), the aggregate trade balance is equal to the sum of net imports,

\begin{equation}
D_i=\sum_{s=1}^S\left(
\sum_{j=1}^J\frac{p_{ij}^sy_{ij}^s}{\tau_{ij}^s}
-
\sum_{j=1}^J\frac{p_{ji}^sy_{ji}^s}{\tau_{ji}^s}\right),
\label{eq:tradebalance}
\end{equation}
where bilateral trade imbalances are endogenously determined within the model.  Note the country indices in the second term are reversed, and represents exports from country  \( i \) to country \( j \).  A value of \( D_i > 0 \) indicates net exports are negative, and indicates a trade deficit for country  \( i \).

\subsubsection*{Goods Market Clearing}

The total supply of goods is consumption plus the intermediate inputs used in each sector,
\begin{equation}
\sum _{s=1}^{S} \sum _{i=1}^{J} \left( x_{ij}^{sk} \right)+c_{j}^{k}=y_{j}^{k}.
 \label{eq:goodmarket}
\end{equation}

\subsubsection*{Labor Market Clearing}
Finally, labor markets must clear in each country,
\begin{equation}
 \sum _{s=1}^{S} {\sum _{i=1}^{J}  {l_{ij}^{s}}}  =L_{j}.
 \label{eq:LaborMkt}
\end{equation}

\section{Calibration and Optimal Tariff Setting}
\label{sec:calibration}

\subsection{Data on Flows and Tariffs}

We use the 2016 Revision of the World Input-Output Database (WIOD), as described by \cite{timmerIllustratedUserGuide2015} and \cite{timmerAnatomyGlobalTrade2016}, for data on industry-level trade flows and input-output flows.  This dataset includes origin-destination-sector expenditures,  \( p_{ij}^{s}y_{ij}^{s}/\tau_{ij}^{s}  \), net of tariffs, and sector-to-sector input-output flows,  \( p_{j}^{k}x_{j}^{sk} \).  Note that in our model,  \( p_{ij}^{s}y_{ij}^{s} \) \ is expenditures from the consumer side, which include tariffs.  The flows in the WIOD data do not include tariffs, therefore we must add them to obtain tariff-inclusive expenditures.  We use data on tariff rates from the United Nations' Trade Analysis Information System (TRAINS) database.  We use effectively applied weighted average tariff rates as our measure for baseline tariffs, \( \tau_{ij}^{s} \).  Table \ref{tbl:data_sector_taus_table} reports pre-trade war tariff rates for each of our tradable goods-producing industries.\footnote{Construction is included the Goods-Producing Industries super-sector classification by the United States' Bureau of Labor Statistics.  We exclude construction from our grouping of tradable goods-producing industries, and include it as a service industry.  Service industries are still tradable, but they are not subject to tariffs.}  The column "USA" reports tariff rates enacted by the United States for imports from China, and the column "CHN" reports tariff rates enacted by China for imports from the United States.\footnote{The tariff data provided by TRAINS is available at the 2-digit International Standard Industrial Classification of All Economic Activities (ISIC) Rev 3 level, which we convert to match the ISIC Rev. 4 sectors in the WIOD using the concordance provided by \citet{pahlJobsProductivityGrowth2022}.}
\begin{table}[htb]
\caption{Tariff Rates between USA and China by Industry (Pre-Trade War) }
\label{tbl:data_sector_taus_table}
\begin{center}
\begin{tabular}{cclSS}
\toprule
Industry&ISIC Codes&Description&{USA}&{CHN}\\ \midrule
1&A01&Crop \& Animal Production&2.30&11.29\\
2&A02&Forestry&2.19&9.43\\
3&A03&Fishing \& Aquaculture&0.54&8.14\\
4&B&Mining&0.40&0.37\\
5&C10-C12&Food, Beverages \& Tobacco&4.94&16.20\\
6&C13-C15&Textiles, Leather \& Apparel&8.88&11.67\\
7&C16&Wood Products&2.25&4.92\\
8&C17&Paper Products&0.00&5.64\\
9&C18&Printing \& Recorded Media&0.10&3.71\\
10&C19&Coke \& Refined Petroleum&0.93&4.91\\
11&C20&Chemicals&3.24&7.27\\
12&C21&Pharmaceuticals&1.25&4.84\\
13&C22&Rubber \& Plastics&3.32&9.95\\
14&C23&Non-Metallic Mineral Products&3.40&11.56\\
15&C24&Basic Metals&1.39&5.21\\
16&C25&Fabricated Metals&2.53&10.41\\
17&C26&Electronics \& Optics&0.98&12.74\\
18&C27&Electrical Equipment&2.43&9.03\\
19&C28&Machinery&1.33&8.63\\
20&C29&Motor Vehicles&2.50&13.47\\
21&C30&Other Transport&1.91&7.77\\
22&C31--C32&Furniture&2.84&14.10\\
\midrule  & $\overline{\tau}$ & Weighted Average Tariff  & 2.97 & 9.95 \\ \bottomrule
\end{tabular}
\end{center}
\end{table}

\subsection{Calibration}
\label{subsec:calibration}
We calibrate our model to fit exactly the data in the initial period.\footnote{See \citet{dingelSpatialEconomicsGranular2020} for a discussion of counterfactual analyses in quantitative models using the calibrated share form approach, where parameters are chosen to match the data exactly.}  First we normalize wages,  \( w_{j}=1 \), and sectoral productivities,  \( z_{j}^{s}=1 \), for all countries and sectors.  We are interested in changes from baseline and these initial normalizations do not affect our results in terms of the counterfactual impact of changes in tariff rates.

\subsubsection*{Sectoral Elasticities}

We take estimates of the sectoral elasticities, \(  \sigma _{s} \), from \citet{fontagneTariffbasedProductlevelTrade2022a}, who provide sectoral elasticity estimates for each of the tradable WIOD sectors.  Their estimates are based on product-level variation in tariff rates and trade-flows, which they pool by sector to estimate sectoral elasticities.  They are unable to estimate elasticities for service sectors, because trade in service sectors is not subject to tariffs.  We set elasticities for service industries equal to 5.5, which is near the median value of their product-level trade elasticities.%

\subsubsection*{Aggregates} %

We calibrate the aggregate trade balance directly from equation (\ref{eq:tradebalance}) and initial tariff revenues from equation (\ref{eq:tariffrevenue}).  Using equation (\ref{BC}), we compute income, ${I}_i$, as
\begin{equation} 
{I}_i=\sum_{s=1}^S p_{i}^{s}c_{i}^{s}=\sum _{s=1}^{S} \sum _{i=1}^{J} p_{ij}^{s}y_{ij}^s-\sum_{s=1}^{S}p_i^sx_i^s,
\end{equation}
 and then re-write equation (\ref{BC2}) to calibrate the aggregate labor supply to
 \begin{equation} 
L_i=\frac{{I}_i-T_i-D_i}{w_i}.
\end{equation}

\subsubsection*{Prices and Quantities}%
Due to the production function for differentiated goods \eqref{eq:yij_def} being constant returns to scale, the shares of intermediate inputs used in the production of differentiated output for a given destination and sector will be proportional to that destination's share of sectoral expenditures, net of tariffs and iceberg trade costs,
\begin{equation*}
    \frac{p_{j}^{k}x_{ij}^{sk}}{p_{j}^{k}x_{j}^{sk}}=\frac{\frac{p_{ij}^{s}y_{ij}^{s}}{ t_{ij}^{s}\tau_{ij}^{s}}}{ \sum _{i}^{} \left( \frac{p_{ij}^{s}y_{ij}^{s}}{ t_{ij}^{s}\tau_{ij}^{s}} \right) }.
\end{equation*}
This equation allows us to recover $p_{j}^{k}x_{ij}^{sk}$, where the other components are taken directly from the data on trade flows and input output flows.  This allows us to calibrate sectoral labor allocations as
\begin{equation} 
     l_{ij}^{s}=\frac{1}{w_{j}} \left( \frac{p_{ij}^{s}y_{ij}^{s}}{ \tau_{ij}^{s}}- \sum _{k=1}^{S}p_{j}^{k}x_{ij}^{sk} \right), 
    \end{equation}
which provide differentiated output, 
\begin{equation} 
     y_{ij}^{s}=\frac{z_{j}^{s}}{t_{ij}^{s}}l_{ij}^{s}, 
     \label{eq:yij_calibration}
    \end{equation} 
assuming one knows $t_{ij}^{s}$, which we discuss below.  We back out prices by dividing trade flows by prices,
\begin{equation} 
    p_{ij}^{s}=\frac{p_{ij}^{s}y_{ij}^{s}}{y_{ij}^{s}}. 
    \label{eq:pij_calibration}
   \end{equation} 
We then compute taste shares, $\gamma_{ij}^s$, as
\begin{equation} 
\gamma_{ij}^s=\frac{p_{ij}^s(y_{ij}^s)^{\frac{1}{\sigma_s}}}{\sum_{j=1}^Jp_{ij}^s(y_{ij}^s)^{\frac{1}{\sigma_s}}}.
\label{eq:gamma_calibration}
\end{equation}
Lastly we calibrate sector shares as
\begin{equation} 
a_i^s=\frac{p_i^sc_i^s}{{I}_i}.
\end{equation}
We use prices \eqref{eq:pij_calibration} and taste-shares \eqref{eq:gamma_calibration} to get the price index, $p_i^s$, for each sector from equation \eqref{eq:pi_def}, and then compute $y_i^s$ from equation \eqref{eq:yij_def}.

If we have data on iceberg trade costs and other non-tariff trade barriers, we can include it directly in our calibration through equation \eqref{eq:yij_calibration}, which then allows us to uniquely pin down $\gamma_{ij}^s$.  If, however, we lack data on iceberg trade costs, then we need to make assumptions about the relative contribution of trade costs and preferences in explaining the bilateral trade flow asymmetries we observe in the data.  Most papers assume either all variation in this equation is due to preferences or that all variation is due to trade costs.  For now, we assume that $t_{ij}^{s}=1$, i.e. that there are no unobserved iceberg trade costs, which means that imbalances arise solely due to asymmetries in preferences.  In Appendix \ref{subsec:iceberg} we explore the sensitivity of our results to this assumption, by presenting an alternative calibration method that allows for unobserved iceberg trade costs, and show that our results hold regardless of the exact source of bilateral imbalances.    

\subsubsection*{IO Shares} %

From fixed proportions in the utility function we have, $\alpha _{j}^{sk}y_{ij}^{s} = x_{ij}^{sk} $.  Multiplying and dividing by $p_{j}^{k}$, gives the expression
 \[  \alpha _{j}^{sk}y_{ij}^{s}= \left( \frac{p_{j}^{k}x_{ij}^{sk}}{p_{j}^{k}} \right).  \] 
Summing over destinations, \( i \), using intermediate input market clearing, and rearranging gives each sectors final goods consumption shares as,
 \[a_{j}^{sk}=\frac{p_{j}^{k}x_{j}^{sk}}{  p_{j}^{k} \sum _{i}^{}  y_{ij}^{s} }, \] 
where $p_{j}^{k}x_{j}^{sk}$ are the flows reported in the input-output tables, and differentiated output quantitites, \( y_{ij}^{s} \), and sector final output prices, $p_{j}^{k}$, were calibrated above.

 \subsection{Computation of Optimal Tariffs}
\label{sec:optimal_tariffs_algorithm}
 We compute optimal tariff rates by numerically finding tariffs that maximize the welfare of households in that country.   Following calibration, we are able to evaluate the impact of a counterfactual change in tariffs by finding prices and allocations that satisfy household and firm optimization conditions and obey market clearing.  For any arbitrary change in exogenous parameters, including tariff rates, computing the counterfactual equilibrium involves finding wages such that labor market clearing holds in each country, conditional on all the equations laid out in Section \ref{sec:model} also holding.  For a candidate counterfactual set of wages, we update prices using \eqref{eq:pi_def}  and \eqref{eq:pij_sol}.  Given these prices, we find output from \eqref{eq:yij_def}, and use \eqref{eq:lij_sol} to update labor and intermediate inputs allocations, respectively.  From those, we are able to update our aggregates using equations \eqref{eq:tariffrevenue}--\eqref{eq:LaborMkt}.  We then use a nonlinear solver to find wages such that the aggregate labor allocations implied by \eqref{eq:LaborMkt} are equal to the labor endowment in each country.
 
 The application motivating our analysis is a tariff war between two countries with a bilateral trade imbalance, therefore, we focus our analysis on optimal tariff setting between two trading partners, \( i \) and \( j \). To compute optimal tariffs, we solve the problem of a government choosing tariffs with a specific trading partner to maximize domestic welfare.  Specifically, the government for country \( i \) chooses tariffs in each sector, \( \boldsymbol{\tau}_{i} \equiv \{\tau_{ij}^{s}\}_{s=1}^S \), to solve
\begin{equation}
    \boldsymbol{\tau}_{i}^{*} = \underset{\{\tau_{ij}^{s}\}_{s=1}^S}{\arg\max}\ W_i \equiv \underset{\{\tau_{ij}^{s}\}_{s=1}^S}{\arg\max}\ \prod _{s=1}^{S} (c_{i}^{s})^{a_{i}^{s}}
\label{eq:optimal_tariffs}
\end{equation}
subject to the constraint that \( \tau_{ij}^{s} \geq 1\) and, for specifications where services are included in the model, that tariffs are set equal to 1 for service sectors.  Optimal unilateral tariffs are given by the solution to this problem under the assumption that all partner countries will leave their tariffs unchanged.  Optimal Nash Equilibrium tariffs are given by the set of tariffs that solve the above problem for each country, taking as given their partner is setting tariffs equal to their Nash tariff.  

In general, there is no analytic solution for best response tariffs or Nash tariffs.  Consequently, standard solution methods require evaluating counterfactual welfare for potential tariff candidates to find the candidates that numerically yield the highest welfare, which necessitates solving the model a large number of times.  Past quantitative work has dealt with this computational intensity by focusing on small models, for example with one or two sectors (see \citeauthor{heArmingtonAssumptionSize2017a}, \citeyear{heArmingtonAssumptionSize2017a}; or \citeauthor{chattopadhyayNashEquilibriumTariffs2019}, \citeyear{chattopadhyayNashEquilibriumTariffs2019}), by focusing on environments where optimal tariff setting is either shown to be, or restricted to be, uniform across sectors \citep{caliendoSecondbestArgumentLow2023}; or by characterizing properties of optimal tariffs without calculating specific values \citep{costinotMicroMacroOptimal2020a}.  The general approach to the problem of numerically solving for optimal tariff rates is given by Algorithm \ref{alg:br_tau} (A more detailed approach is described by Algorithm \ref{alg:genetic} in Appendix \ref{subsec:ga}).  Running Algorithm \ref{alg:br_tau} a single time with baseline tariffs yields optimal unilateral tariffs.

\begin{algorithm}
    \caption{Computation of Best Response Tariff Rates}
    \label{alg:br_tau}
    \begin{algorithmic}[1]
    \renewcommand{\algorithmicrequire}{\textbf{Setup:}}
    \renewcommand{\algorithmicensure}{\textbf{Algorithm:}}
    \REQUIRE Calibrate model according to Section \ref{subsec:calibration}
    \ENSURE  \textit{numerical solution to \eqref{eq:optimal_tariffs}}%
    \\ \textit{Optional Initialization Step : Seed population with initial tariff candidates}
     \WHILE{ \( \max{ (\lVert \Delta  \boldsymbol{\tau}_{i}^{*} \rVert,\lVert\Delta W_{i}^{*}\rVert) } \geq \textit{tol} \)} 
     \STATE Generate population of tariff candidates, $\boldsymbol{\tau}_{i}$, including any seeds \label{alg:br_tau:pop_step}
     \STATE For each tariff candidate, $\boldsymbol{\tau}_{i}$ evaluate counterfactual welfare $W_{i}$
     \STATE Select top performing candidate as $\boldsymbol{\tau}_{i}^{*}$, with corresponding welfare $W_{i}^{*}$.  Include $ \boldsymbol{\tau}_{i}^{*}$ as seed in next iteration. 
    \ENDWHILE
    \RETURN $ \boldsymbol{\tau}_{i}^{*} $ 
    \end{algorithmic} 
    \end{algorithm}

We compute Nash Equilibrium tariffs by sequentially iterating best response tariffs.  That is, we compute optimal tariffs for country  \( i \) taking country  \( j \)'s tariffs as given.  We then compute optimal tariffs for country  \( j \), taking country  \( i \)'s tariffs as given, and equal to the the optimal tariffs we just solved for.  We follow this process iteratively, sequentially solving for each country's optimal tariffs and updating them until the process converges.\footnote{In principle, it should be possible to expand our methodology to optimal tariff setting between arbitrary numbers of countries.  No special adjustment is necessary to compute optimal unilateral tariffs in which a country simultaneously sets tariffs for all partner countries.  Therefore Nash Equilibrium tariffs can still be computed using the strategy of iteratively computing optimal best response tariff rates for each country until no country chooses to deviate.  However, convergence from such an approach may become prohibitively slow as the number of countries involved grows.}  This is a standard method of computationally finding Nash equilibria, see for example \cite{oladiStableTariffsRetaliations2005}, although it is not guaranteed to converge in finite time in all classes of games \citep{Nisan2007}.  We find that iterating best response functions does converge in a reasonable number of steps, therefore we did not explore alternative methods for finding Nash equilibria.   Algorithm \ref{alg:nash_tau} presents the general approach to iterating best response tariffs to find Nash Equilibrium tariffs.

\begin{algorithm}
    \caption{Computation of Nash Equilibrium Tariff Rates}
    \label{alg:nash_tau}
    \begin{algorithmic}[1]
    \renewcommand{\algorithmicrequire}{\textbf{Setup:}}
    \renewcommand{\algorithmicensure}{\textbf{Algorithm:}}
    \REQUIRE Calibrate model according to Section \ref{subsec:calibration}
    \ENSURE  \textit{(Countries are labeled USA and CHN for exposition)}%
    \STATE Set $\boldsymbol{\tau}_{USA}$ and $\boldsymbol{\tau}_{CHN}$ equal to baseline tariffs 
    \\ \textit{When computing optimal tariffs, partner tariffs are taken as given}
     \WHILE{ \( \max{ (\lVert \Delta  \boldsymbol{\tau}_{USA}^{*} \rVert,\lVert\Delta W_{USA}^{*}\rVert,\lVert \Delta  \boldsymbol{\tau}_{CHN}^{*} \rVert,\lVert\Delta W_{CHN}^{*}\rVert) } \geq \textit{tol} \)} 
     \STATE Compute $\boldsymbol{\tau}_{USA}^{*}$ using Algorithm \ref{alg:br_tau} to get $W_{USA}^{*}$
     \STATE Update $\boldsymbol{\tau}_{USA}=\boldsymbol{\tau}_{USA}^{*}$
     \STATE   Compute $\boldsymbol{\tau}_{CHN}^{*}$ using Algorithm \ref{alg:br_tau} to get $W_{CHN}^{*}$
     \STATE Update $\boldsymbol{\tau}_{CHN}=\boldsymbol{\tau}_{CHN}^{*}$
    \ENDWHILE
    \RETURN $ \boldsymbol{\tau}_{USA}^{*}$ \textit{and} $ \boldsymbol{\tau}_{CHN}^{*}$ 
    \end{algorithmic} 
    \end{algorithm}

\citet{huangMultipleNashEquilibria2013} show that, for certain parameterizations, it is possible to have multiple Nash equilibria in a simple two country two good CES framework.  Particularly with low trade elasticities, they demonstrate the existence of Nash equilibria with extremely high tariff rates, including rates greater than 100,000 percent.  For other parameterizations, the Nash equilibrium will be unique.  To limit any potential impact of multiple equilibria, we restrict our analysis to tariff rates capped above at 400 percent.  This rate exceeds the 99th percentile of industry level tariff rates in the TRAINS data, which is approximately 55 percent. This cap is also higher than the maximum reported ad valorem tariff rate of 350 percent set on an 8-digit Harmonized System product code by the United States in both 2014 and 2023, according to the United States International Trade Commission's Tariff Database.\footnote{https://dataweb.usitc.gov/tariff/annual}  We confirm that moderately lowering or raising this upper bound does not affect our quantitative results.

It is important to note that the general approach in Algorithm \ref{alg:br_tau} is separable from the method used to generate the population of candidate solutions.  For example, a traditional local optimization algorithm might have a population size of 1, and update the singular candidate according to an analytic formula.  This is how local root-solving algorithms such as the Newton-Raphson and Midpoint methods work.  There is no guarantee, however, that an arbitrary method for step \ref{alg:br_tau:pop_step} of Algorithm \ref{alg:br_tau} will necessarily converge to the actual solution of \eqref{eq:optimal_tariffs}.  Our experience is that local optimization solvers, in isolation, perform poorly.  With that caveat, there are a number of global optimization heuristics available to researchers, and we expect many of these approaches to be reasonable choices in our context.  Our expectation is that researchers will be best served by choosing an approach that is conservative in terms of trading researcher time for compute time.  In the related literature of hyper-parameter optimization, \cite{liHyperbandNovelBanditbased2018} show that random search with twice the run-time outperforms a number of more sophisticated methods.  Similarly, the general approach of computing Nash equilibrium tariffs using Algorithm \ref{alg:nash_tau} is separable from the exact method used to compute best response tariffs.

We adopt a hybrid genetic algorithm approach for the problem of updating optimal tariff candidates in Algorithm \ref{alg:br_tau}.  Our approach involves finding candidate solutions using a genetic algorithm, which we then refine by running a local optimization algorithm.  In Section \ref{sec:application} we show the welfare function for deviations from Nash equilibrium tariffs is convex, which local optimization algorithms are able to take advantage of when near the solution.  We refer readers to \cite{mitchellIntroductionGeneticAlgorithms1998} for an overview of genetic algorithms and \citet{backOverviewEvolutionaryAlgorithms1993} for an overview of evolutionary algorithms --- of which genetic algorithms are a subclass. \citet{kehoeImprovingTradePolicy} discuss the use of a genetic algorithm for calibrating industry level trade elasticities.  Genetic algorithms are widely used due to several desirable properties, including their ease of modification, ability to escape local optima, that they are readily available to researchers for easy implementation,\footnote{MATLAB includes a modifiable implementation of a genetic algorithm as part of its Global Optimization Toolbox, PyGad offers an open source Python implementation of a genetic algorithm, and user written packages exist for Julia, R, Fortran, and other languages.} and they can be seeded with initial guesses to speed convergence.  The ability to seed candidate solutions is particularly important for accelerating convergence when we iterate best response tariffs to compute Nash equilibrium tariffs.

We discuss our implementation of a hybrid genetic algorithm in Appendix \ref{subsec:ga}.  As noted previously, there is nothing inherently unique about our chosen approach, and feasible alternatives exist, including random search, simulated annealing \citep{wuIntroductionSimulatedAnnealing1998}, particle swarm optimization \citep{kennedyParticleSwarmOptimization1995}, and hyperband optimization \citep{liHyperbandNovelBanditbased2018}.  It should be noted that grid search is implausible due to the computational unfeasibility of evaluating all points on the grid, even for small grids.\footnote{As noted in the introduction, even a discretized solution space of only 10 possible tariff rates for each sector yields 10 Sextillion possible combinations.}  We do not provide a detailed comparison on the relative performances of various global optimization heuristics, in part because each of them can be modified and adapted to perform better on specific problems.  For instance, the genetic algorithm described in Appendix \ref{subsec:ga} will outperform a generic implementation of a genetic algorithm in terms of both accuracy and speed when applied to our problem of computing optimal tariff rates.  

After convergence, we verify through exhaustive search that there are no single-sector deviations from Nash equilibrium tariffs that improve welfare. We evaluate counterfactual welfare changes for all tariffs between 0 and 25 percent to two decimal places for each sector.  This encompasses the range of optimal tariffs we find, and we verify separately with a coarser grid that deviations with higher tariff rates are also not welfare-improving.  This requires solving the model for 110,000 different sets of counterfactual tariffs, which is computationally feasible.

\section{Numerical Exercises}
\label{sec:numerical_exercises}

To better understand the interaction between optimal tariff rates and aggregate and bilateral trade deficits, we conduct several numerical exercises.  For each numerical exercise, we start by constructing a symmetric world, with a single sector, where each country has total expenditures equal to 100.  We start with zero tariffs, and assume expenditures are equally divided among trading partners, and that there are no service sectors or input-output linkages.  To create varying bilateral deficit levels, we add a fixed amount, $d/2$ to exports from Country 2 to Country 1, and subtract the same amount from the reverse flow.   This leads to Country 1 having a trade deficit of $d$ with Country 2, prior to any implementation of tariffs.  We vary $d$ to understand the resulting implications for a range of possible trade deficits.  We calibrate our model to match these expenditure flows according to the methodology laid out in Section \ref{sec:calibration}, and then we compute optimal unilateral and Nash equilibrium tariffs following the procedure in Section \ref{sec:optimal_tariffs_algorithm}.

\subsection{Two Country Numerical Exercises}
\label{subsec:two_country_numerical_exercises}
In a two country framework, bilateral deficits are equal to aggregate deficits.  Aggregate trade deficits enter into the household budget constraint as a transfer.  In a two country framework, this transfer is paid from the country with the trade surplus to the country with a trade deficit.  We consider both the case where a country with a trade surplus is forced to pay the transfer and the case where the country with a trade surplus is allowed to select autarky as an option, in which case no transfer is paid.    We restrict our analysis to deficit levels in which both countries prefer free trade to autarky in the baseline with no tariffs.  Allowing for autarky introduces a novel situation in which a country with a trade deficit may choose to voluntarily restrict their tariff in order to ensure that its trading partner does not choose autarky.  This scenario in our numerical results expands upon the illustrative results in Section \ref{sec:toy_model}, in which we assume that the pattern of trade does not change in response to tariffs.

\begin{figure}[!htp]
\caption{Optimal Tariffs with Varying Trade Imbalances (Two Countries) \label{Fig--Tariff_Rates_2Country} } %
	\begin{minipage}[b]{\linewidth}
		\centering
		 	\includegraphics[scale=0.9]{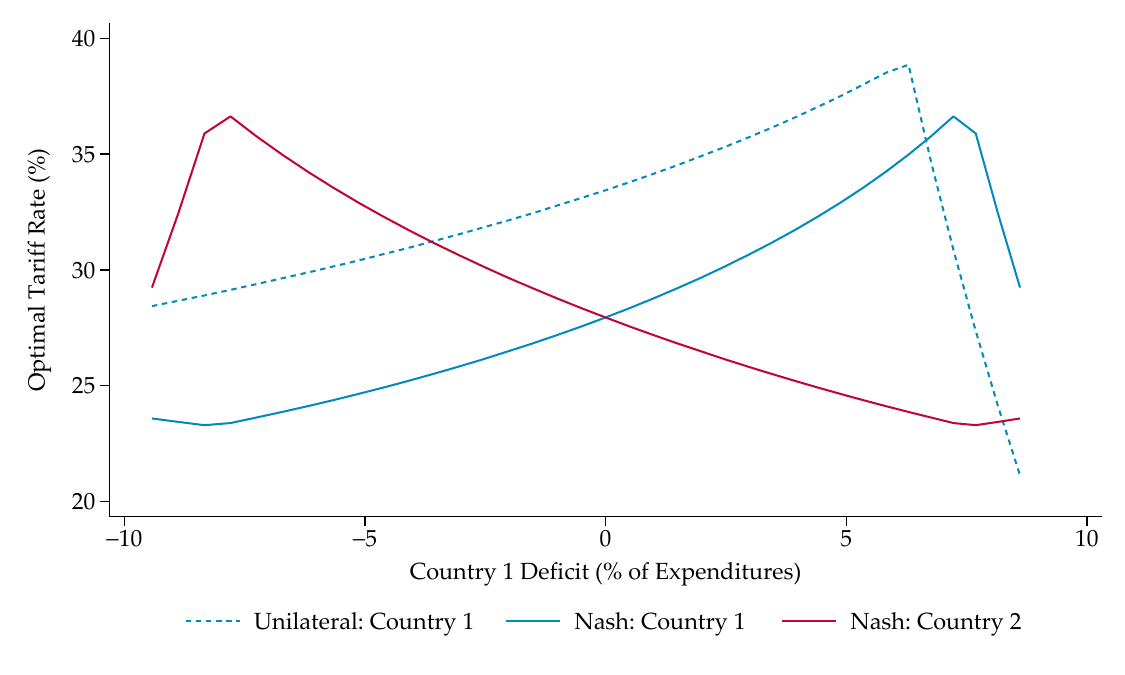}
	\end{minipage}

\end{figure}

Figure \ref{Fig--Tariff_Rates_2Country} illustrates how optimal unilateral tariffs and Nash equilibrium tariffs adjust in response to varying trade deficits. As the trade deficit shifts from negative to positive, the optimal unilateral tariffs set by Country 1 increase steadily, reaching a peak when the trade deficit becomes sufficiently large. This trend reflects Country 1's ability to set tariffs optimally without the constraint of Country 2 potentially opting for autarky. This scenario aligns with \citet{johnsonOptimumTariffsRetaliation1953}, the theoretical analysis in Section \ref{sec:toy_model}, and shows that optimal unilateral tariffs rise consistently with the trade deficit, as long as the trade pattern remains unchanged. Beyond this peak, however, the trade deficit becomes so substantial that Country 1 begins to limit its tariffs to prevent Country 2 from choosing autarky.

Nash equilibrium tariff rates are symmetrical around the origin, as the only difference between the two countries lies in the trade deficit. Focusing on Country 1’s Nash equilibrium tariffs, we observe a pattern similar to that of unilateral tariffs: tariffs rise as the trade deficit grows, then suddenly decline to prevent the trading partner from choosing autarky. Although the logic of the argument is the same as before, Nash tariffs are lower than Unilateral tariffs before reaching their peak. This aligns with our findings in Section \ref{sec:toy_model}, where optimal tariffs are lower when the trading partner also imposes tariffs. In the Nash equilibrium case, the peak arrives later, and tariffs do not fall as sharply. This difference is due to Country 2’s optimal response of setting positive tariffs itself, which makes the threat of autarky less significant.

Our numerical exercises match the familiar result in the literature that optimal unilateral tariffs are positive and lead to welfare gains, for example as discussed by \citet{costinotChapterTradeTheory2014}.   Country 1 experiences welfare gains or losses under the Nash equilibrium tariff rates depending on the value of the deficit, the trade elasticity, and the size of each country.  The Armington elasticity affects both optimal tariff rates and the point at which a trading partner would opt for autarky, with lower Armington elasticities corresponding to greater potential gains from trade wars, and higher Armington elasticities leading to scenarios under which no range of trade deficit would lead to welfare gains under Nash tariffs.  Even under parameterizations where it will not be possible for a country to gain from Nash equilibrium tariffs, the welfare losses from a trade war decrease as the country's trade deficit increases.  

\subsection{Three Country Numerical Exercises}
In a three country framework, countries are able to have bilateral trade deficits even when there are no aggregate trade deficits.  In our numerical exercises, we generate scenarios with varying levels of bilateral trade deficits where each trade deficit (surplus) is perfectly offset by a trade surplus (deficit) with the other trading partner.  Specifically, for a bilateral deficit of $d$ between Country 1 and Country 2, we subtract $d/2$ from Country 3's exports to Country 1 and add $d/2$ to exports from Country 1 to Country 3.  We apply a mirrored adjustment for Country 2's trade flows with Country 3.  This ensures that no country has an aggregate trade imbalance, despite its bilateral imbalances.  We consider optimal tariff setting between Country 1 and Country 2, and assume that Country 3 does not participate and leaves its tariff rates unchanged.

\begin{figure}[!htp]
\caption{Optimal Tariffs and Welfare Changes with Varying Trade Imbalances \\ (Three Countries, Bilateral Deficits Only) \label{Fig--OptimalTariffs_3Country} } %
	\begin{minipage}[b]{\linewidth}
		\centering
		 	\includegraphics[scale=0.9]{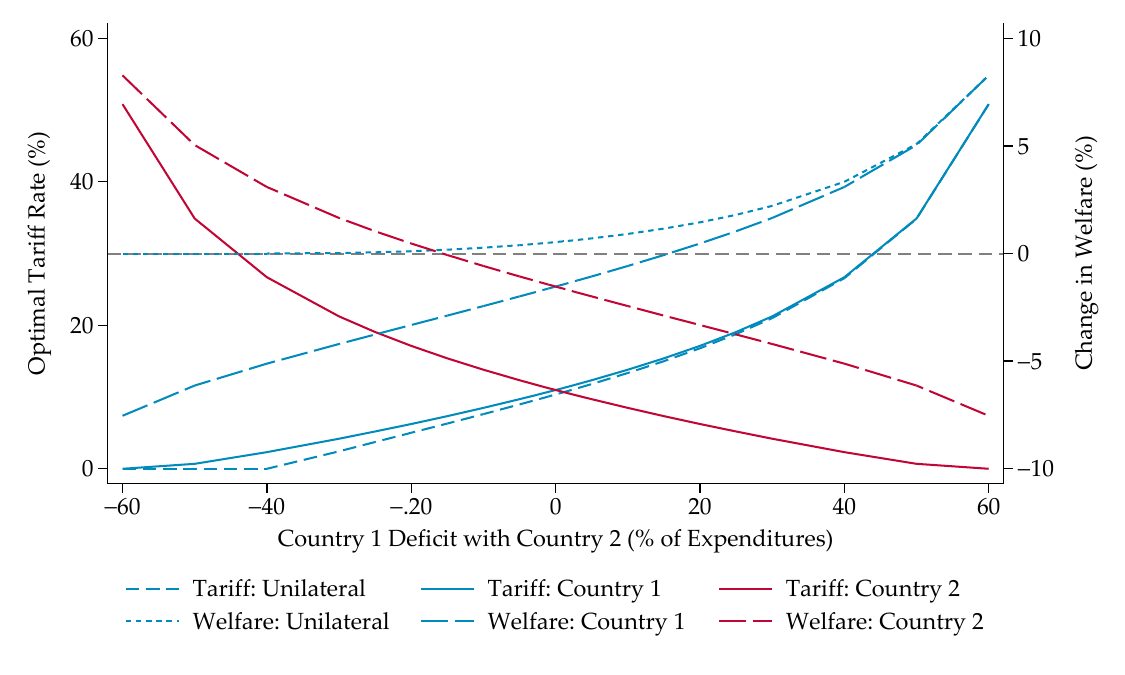}
	\end{minipage}

\end{figure}

Figure \ref{Fig--OptimalTariffs_3Country} shows how optimal tariff rates change as the bilateral trade deficit varies.  When there are no aggregate trade deficits, the motive of a country with a trade surplus to choose autarky to avoid paying a transfer is no longer present.  This means that Country 1's optimal unilateral and Nash equilibrium tariffs increase monotonically as the trade deficit increases.  These figures make it clear that both aggregate and bilateral deficits impact optimal tariffs.  They further highlight that a country with a large bilateral trade deficit might gain economically from instigating a trade war, at the expense of its trading partner.  %

When there are only bilateral deficits, and no aggregate deficits, tariffs are effectively mirrored between Country 1 and Country 2. As Country 1 exhibits a larger bilateral trade deficit with Country 2, its Nash equilibrium tariff rate increases while Country 2's Nash equilibrium tariff rate decreases.  This might seem to suggest that a country with a bilateral deficit will always have higher optimal tariffs than its trading partner; however, this need not be true in general.  Optimal tariff rates interact not only with the bilateral imbalance between partners, but also bilateral and aggregate imbalances with further trading partners.

\begin{figure}[!htp]
\caption{Optimal Tariffs with Varying Trade Imbalances \\ (Aggregate and Bilateral Deficits with Multiple Countries) \label{Fig--Tariff_Rates_3Country_AggImb} } %
	\begin{minipage}[b]{\linewidth}
		\centering
		 	\includegraphics[scale=0.9]{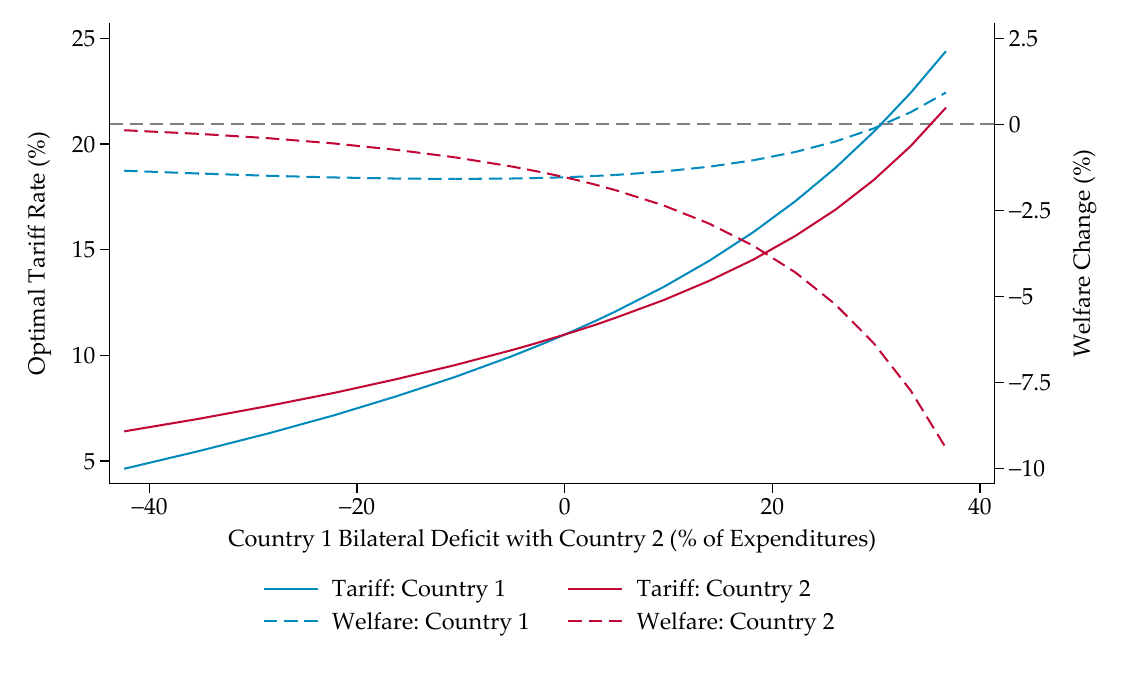}
	\end{minipage}

\end{figure}

Figure \ref{Fig--Tariff_Rates_3Country_AggImb} presents the Nash equilibrium tariffs and resulting welfare changes under the scenario where Country 1 has a trade deficit with Country 3, and Country 2 has a trade surplus with Country 3.  Instead of adjusting trade with Country 3 by $d/2$, we adjust it in the opposite direction by a fixed amount.\footnote{For this scenario, we set the deficit between Country 1 and Country 3 to 15, and the same for the surplus between Country 2 and Country 3.  Recall we set income, net of deficits, equal to 100 for these examples.}  Therefore, Country 1 has an aggregate trade deficit greater than its bilateral trade deficit with Country 2.  Under this scenario, optimal tariffs are no longer mirrored; and in fact, Nash equilibrium tariffs for Country 2 increase as the bilateral deficit increases between Country 1 and Country 2, until they are higher than Country 1's tariffs.  This scenario more closely mirrors the economic situation between the United States and China, where the United States maintains a deficit with the rest of the world and China maintains a surplus.  This helps explain why, under several model specifications, we will find that China's Nash equilibrium tariffs are higher than those of the United States, despite China having a trade surplus with the United States.  It continues to be true, however, that a larger trade deficit increases the welfare of the country with the deficit in a trade war, and lowers the welfare of the country with the surplus.

\begin{figure}[htp]
\caption{Welfare Impact of a 10 Percent Tariff with Varying Trade Imbalances \label{Fig--FakeData_FlatTariff} } %
	\begin{minipage}[b]{\linewidth}
		\centering
		 	\includegraphics[scale=0.9]{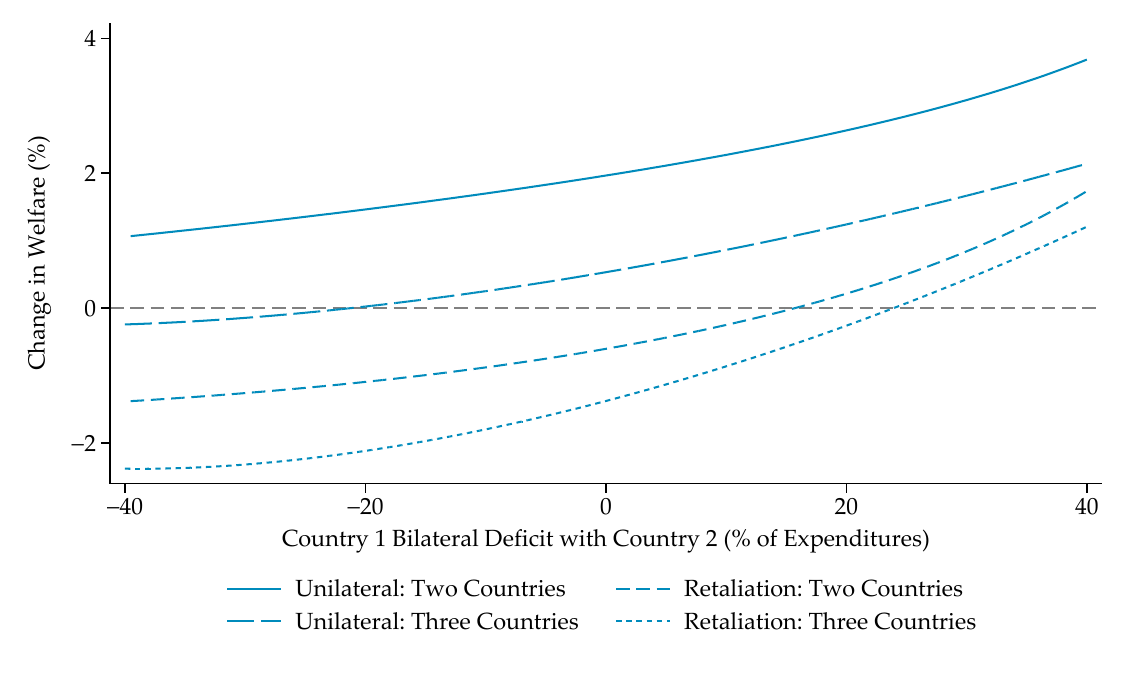}
	\end{minipage}

\end{figure}

Although our focus is on optimal tariffs rates, it is important to note that trade imbalances do not just affect optimal tariff rates.  Trade imbalances also alter the welfare implications of arbitrary tariffs.  Figure \ref{Fig--FakeData_FlatTariff} highlights this point by showing how the welfare of Country 1 changes as its trade balance varies when Country 1 places a flat 10 percent tariff on imports from Country 2.  We consider both the case where Country 1 places unilateral tariff and where Country 2 retaliates by putting a flat 10 percent tariff on imports from Country 1.  We find that Country 1's welfare increases monotonically as its bilateral trade deficit grows (or its bilateral trade surplus shrinks), for both unilateral and retaliatory 10 percent tariffs.\footnote{We ignore the possibility of autarky in Figure \ref{Fig--FakeData_FlatTariff}, which we discuss in Section \ref{subsec:two_country_numerical_exercises}, and here focus on the scenario where the trade imbalances cannot be strategically avoided.}  We show these results hold in both the scenario with two countries, where bilateral imbalances are aggregate imbalances, and with three countries, where bilateral imbalances can exist with or without aggregate imbalances.

\section{Application to U.S.--China Trade War}
\label{sec:application}
In this section, we apply our methodology to study a trade war between the United States and China. Our framework is particularly relevant for understanding a trade conflict between these two countries, given their mutual importance to each other as major trading partners and their unparalleled trade imbalance.

Table \ref{tbl:Optimal_Taus_Full} presents optimal unilateral and Nash equilibrium tariffs for each of the 22 tradable goods-producing sectors (unilateral tariffs are reported only for the United States).  Results are reported for changes relative to two distinct baseline economies.  The first scenario is when the model is calibrated to match exactly the 2014 baseline economy.  This baseline was chosen because it is the most recent year for which we have input-output data available, and because it is two years prior to the initial election of President Trump, who used the executive powers granted through his office to initiate the first round of tariff increases for U.S. imports of goods from China.  The second baseline we consider is a world economy with free trade between all bilateral country pairs, including the United States and China.  We construct the free trade baseline by calibrating the model to match 2014 data and computing the resulting counterfactual world economy after setting all tariffs equal to 0 ($\tau^s_{ij}=1$) for all country pairs $ij$ and sectors $s$.  For each baseline, we solve for optimal unilateral and Nash equilibrium tariffs between the United States and China.  Optimal tariffs are non-uniform because sectoral elasticities vary across sectors and, in the pre-trade war baseline, partner tariffs vary by sector.  The table also reports the trade-weighted average tariff rate, $\overline{\tau^{*}}$, for each baseline, where the weights for each sector are equal to trade values in the baseline data.

\begin{table}[!htb]
\caption{Optimal Tariffs Between United States and China\\(Model with Services and Input-Output Linkages)}
\label{tbl:Optimal_Taus_Full}
\begin{center}
\begin{tabular}{cSSSSSS}
\toprule
& \multicolumn{3}{c}{Pre-Trade War Baseline} & \multicolumn{3}{c}{Free Trade Baseline} \\
\cmidrule(lr){2-4} \cmidrule(lr){5-7}
 & {Unilateral} & \multicolumn{2}{c}{Nash Tariffs} & {Unilateral} & \multicolumn{2}{c}{Nash Tariffs} \\
  \cmidrule(lr){2-2} \cmidrule(lr){3-4} \cmidrule(lr){5-5} \cmidrule(lr){6-7}
{Sector} & {USA} & {USA} & {CHN} & {USA} & {USA} & {CHN} \\
\midrule
1 & 15.54 & 16.12 & 20.25 &  16.65 & 16.85 & 19.19 \\ 
2 & 14.00 & 13.49 & 18.16 &  16.25 & 23.25 & 15.01 \\ 
3 & 11.30 & 11.27 & 18.62 &  9.89 & 13.40 & 18.89 \\ 
4 & 11.01 & 11.01 & 14.73 &  10.38 & 10.62 & 13.61 \\ 
5 & 13.86 & 13.99 & 18.14 &  13.55 & 13.84 & 16.61 \\ 
6 & 10.33 & 10.33 & 20.34 &  6.35 & 6.44 & 19.06 \\ 
7 & 12.15 & 12.33 & 16.98 &  11.78 & 12.03 & 16.39 \\ 
8 & 12.45 & 12.43 & 16.74 &  11.62 & 12.08 & 14.87 \\ 
9 & 13.96 & 13.24 & 18.47 &  12.71 & 12.86 & 14.66 \\ 
10 & 11.13 & 11.16 & 13.91 &  10.23 & 10.39 & 13.15 \\ 
11 & 11.45 & 11.47 & 16.79 &  10.69 & 10.72 & 14.08 \\ 
12 & 11.29 & 11.11 & 15.81 &  10.26 & 11.20 & 16.04 \\ 
13 & 11.47 & 11.50 & 17.37 &  10.86 & 10.89 & 15.33 \\ 
14 & 12.10 & 12.25 & 16.35 &  11.35 & 11.66 & 14.81 \\ 
15 & 8.25 & 8.35 & 15.61 &  8.55 & 7.87 & 13.57 \\ 
16 & 11.39 & 11.42 & 16.57 &  11.07 & 11.13 & 14.56 \\ 
17 & 9.99 & 10.06 & 18.59 &  9.27 & 9.44 & 13.57 \\ 
18 & 8.60 & 8.70 & 16.04 &  7.88 & 7.99 & 13.65 \\ 
19 & 9.82 & 9.96 & 15.24 &  9.22 & 9.42 & 12.75 \\ 
20 & 10.14 & 10.19 & 16.00 &  9.38 & 9.41 & 14.29 \\ 
21 & 12.56 & 12.99 & 16.44 &  12.63 & 12.52 & 14.47 \\ 
22 & 11.40 & 11.52 & 15.72 &  10.74 & 10.91 & 10.08 \\ 
 \midrule  {$\overline{\tau^{*}}$} & 10.34 & 10.41 & 17.04 & 9.17 & 9.30 & 14.74 \\ 
\cmidrule(lr){1-1} \cmidrule(lr){2-2} \cmidrule(lr){3-4} \cmidrule(lr){5-5} \cmidrule(lr){6-7}
 {$\Delta W(\%)$} & 0.025 & 0.008 & -0.138 & 0.039 & -0.003 & -0.132 \\ 
\bottomrule
\end{tabular}
\end{center}
\end{table}

For each counterfactual change in tariffs, we report the corresponding counterfactual change in welfare, which we measure in terms of consumption equivalence.  We report consumption equivalence in percentage change, i.e. \( 100*( \beta_{i}-1) \), where \( \beta_{i} \) is the scalar such that,  
\begin{equation}
\label{eq:cons_equiv}
 \prod _{s=1}^{S} (\beta_{i} c_{i}^{s})^{a_{i}^{s}}  =  \prod _{s=1}^{S} ( c_{i}^{s'})^{a_{i}^{s}} 
\end{equation}
where \( c_{i}^{s} \) is country \(i\)'s consumption of sector \(s\) in the baseline, and \( c_{i}^{s'} \) is its consumption of that sector in the counterfactual equilibrium.  Relative to a free trade baseline, both countries experience welfare losses, with China experiencing larger losses.

\begin{figure}[htp]
\caption{Welfare Consequences of Deviations from Nash Tariffs \\ (Model with Services, IO-Linkages, and Free Trade) \label{Fig--Component_NashVerify} } %
	\begin{minipage}[b]{\linewidth}
		\centering
		 	\includegraphics[scale=0.9]{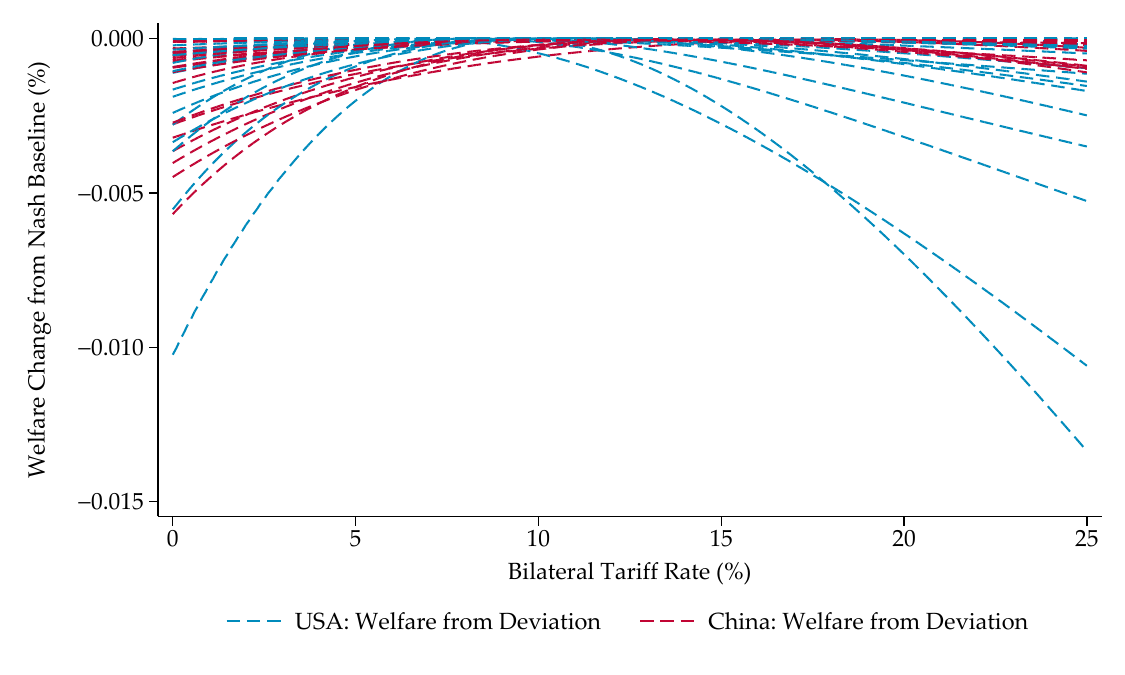}
	\end{minipage}

\end{figure}

We verify that at the Nash equilibrium tariff rates, neither country benefits from individually adjusting the tariff rate in any one sector.  Figure \ref{Fig--Component_NashVerify} plots the welfare consequences of sector-wise deviations from the Nash equilibrium tariffs reported in Table \ref{tbl:Optimal_Taus_Full} under the free trade baseline.  As the figure shows, there are no sectors for which either country could receive higher welfare by unilaterally deviating from the Nash equilibrium tariff rate for that sector.  This exercise provides confidence that our genetic algorithm correctly converges to the Nash equilibrium tariff rates.

\begin{table}[!htb]
\caption{Optimal Tariffs Between United States and China\\(Model without Bilateral or Aggregate Trade Deficits)}
\label{tbl:Optimal_Taus_Full_NoDeficits}
\begin{center}
\begin{tabular}{cSSSSSS}
\toprule
& \multicolumn{3}{c}{Pre-Trade War Baseline} & \multicolumn{3}{c}{Free Trade Baseline} \\
\cmidrule(lr){2-4} \cmidrule(lr){5-7}
 & {Unilateral} & \multicolumn{2}{c}{Nash Tariffs} & {Unilateral} & \multicolumn{2}{c}{Nash Tariffs} \\
  \cmidrule(lr){2-2} \cmidrule(lr){3-4} \cmidrule(lr){5-5} \cmidrule(lr){6-7}
{Sector} & {USA} & {USA} & {CHN} & {USA} & {USA} & {CHN} \\
\midrule
{1} & 13.19 & 14.73 & 15.77 &  2.80 & 15.89 & 14.32 \\ 
{2} & 20.35 & 15.50 & 15.04 &  0.59 & 16.63 & 12.13 \\ 
{3} & 14.44 & 10.02 & 14.03 &  0.41 & 10.03 & 12.50 \\ 
{4} & 9.18 & 9.13 & 9.42 &  7.38 & 9.10 & 8.26 \\ 
{5} & 10.91 & 11.57 & 13.83 &  11.09 & 11.42 & 11.93 \\ 
{6} & 8.83 & 9.06 & 15.96 &  5.07 & 5.44 & 14.49 \\ 
{7} & 10.47 & 10.40 & 12.95 &  10.44 & 10.29 & 11.35 \\ 
{8} & 10.05 & 10.40 & 12.73 &  10.83 & 10.38 & 10.58 \\ 
{9} & 10.34 & 11.59 & 12.96 &  0.75 & 11.36 & 11.43 \\ 
{10} & 9.02 & 9.53 & 10.31 &  7.91 & 9.40 & 8.74 \\ 
{11} & 9.60 & 10.09 & 12.29 &  9.24 & 9.68 & 9.89 \\ 
{12} & 9.54 & 9.84 & 12.43 &  9.04 & 9.75 & 11.21 \\ 
{13} & 9.41 & 9.89 & 13.09 &  9.19 & 9.63 & 10.84 \\ 
{14} & 9.55 & 10.28 & 12.48 &  9.21 & 10.02 & 10.87 \\ 
{15} & 7.23 & 7.75 & 10.98 &  6.85 & 7.54 & 8.94 \\ 
{16} & 9.96 & 10.27 & 12.24 &  9.52 & 10.13 & 10.31 \\ 
{17} & 8.69 & 8.76 & 14.35 &  8.05 & 8.40 & 9.76 \\ 
{18} & 7.22 & 7.47 & 12.01 &  6.88 & 7.11 & 9.42 \\ 
{19} & 8.13 & 8.38 & 11.19 &  8.01 & 8.18 & 8.68 \\ 
{20} & 7.60 & 8.70 & 12.30 &  7.70 & 8.23 & 10.21 \\ 
{21} & 10.90 & 11.18 & 12.29 &  11.20 & 11.11 & 10.42 \\ 
{22} & 9.00 & 9.45 & 11.84 &  8.96 & 9.16 & 5.86 \\ 
 \midrule \rule{0pt}{2.5ex}  {$\overline{\tau^{*}}$} & 8.75 & 8.99 & 12.85 & 7.77 & 8.16 & 10.44 \\ 
\cmidrule(lr){1-1} \cmidrule(lr){2-2} \cmidrule(lr){3-4} \cmidrule(lr){5-5} \cmidrule(lr){6-7}
 \rule{0pt}{2.5ex}  {$\Delta W(\%)$} & 0.015 & -0.019 & -0.065 & 0.027 & -0.059 & -0.037 \\ 
\bottomrule
\end{tabular}
\end{center}
\end{table}
We now explore the importance of bilateral and aggregate trade imbalances for our results.  We eliminate aggregate trade deficits for all countries by calibrating the model to the data, setting $D_{i}=0$ for all countries, and then solving the counterfactual equilibrium and using this as our baseline economy for computing optimal tariffs and welfare changes.\footnote{Changing $D_{i}$ to eliminate trade deficits is the approach followed by \citet{dekleUnbalancedTrade2007} and \citet{caliendoEstimatesTradeWelfare2015}. In Appendix \ref{subsec:no_agg_imbalances} we show that the results are similar if we eliminate aggregate trade imbalances for all countries by scaling trade flows with the rest of the world prior to calibrating the model.}  We eliminate bilateral trade deficits by scaling taste share parameters for U.S. exports to China in traded sectors, $\gamma_{CHN,USA}^{s}$, by a constant proportion, $\zeta$, until the resulting counterfactual equilibrium has no bilateral deficit between the two countries.\footnote{We simultaneously scale other taste shares by a constant proportion $\eta$, so we have $\sum_{j=1,{j \ne USA}}^{J}{\eta\gamma_{CHN,j}^{s}}+\zeta\gamma_{CHN,USA}^{s} =1$.  We also reduce the aggregate trade deficit (surplus) for the United States (China) by the amount of the initial bilateral U.S.-China trade deficit (surplus).  It is important to calculate the taste scaling parameter to eliminate bilateral trade imbalances after making other adjustments.  Otherwise, eliminating tariffs or aggregate trade imbalances may re-introduce bilateral imbalances.  The scaling parameter, $\zeta$, ranges from 1.22 for the pre-trade war baseline economy to 1.14 for the baseline economy with free trade and aggregate trade deficits also eliminated.}       

Table \ref{tbl:Optimal_Taus_Full_NoDeficits} shows results with no aggregate trade deficits and no bilateral trade deficits between the United States and China.  In this scenario, optimal tariffs are lower than our results with bilateral and aggregate trade imbalances, and welfare is lower under Nash Equilibrium tariffs for both the United States and China compared to their respective baseline economies.  These results show that eliminating trade imbalances leads to starkly different welfare implications, with the United States experiencing a welfare loss in such a scenario.  In particular, the welfare change from Nash equilibrium tariffs relative to the existing tariff rate baseline falls from $+0.008$ in \ref{tbl:Optimal_Taus_Full} to $-0.019$ after eliminating bilateral and aggregate trade imbalances.

Table \ref{tbl:Multisector_Full_Deficit_results} shows how our results change across different specifications when we independently eliminate aggregate and bilateral trade imbalances.  We report trade-weighted average tariff rates, where the weights are the pre-trade war trade values in the data.  These results suggest that for the trade war between the United States and China, the key driver of our results in Table \ref{tbl:Optimal_Taus_Full} is the bilateral trade imbalance between the United States and China.  These results highlight how trade imbalances lead to higher optimal tariff rates, consistent with the the numerical examples in Section \ref{sec:numerical_exercises}.  In Appendix \ref{subsec:NumElast}, we calculate the numerical elasticities for each sector for the United States and China and showing how bilateral trade imbalances cause the US to have relatively more inelastic import demand.  This highlights that the intuition from the illustrative model from Section \ref{sec:toy_model} continues to apply in our quantitative framework.

\begin{table}[htb]
\caption{Optimal Tariffs and Welfare Changes (With and Without Deficits)}
\label{tbl:Multisector_Full_Deficit_results}
\begin{center}
\begin{tabular}{cccSS|SSSS}
\toprule
\multicolumn{3}{c}{}& \multicolumn{2}{c}{Unilateral Tariffs} & \multicolumn{4}{c}{Nash Tariffs} \\
\cmidrule(lr){4-5} \cmidrule(lr){6-9} 
& &  & \multicolumn{2}{c}{USA} & \multicolumn{2}{c}{USA} & \multicolumn{2}{c}{CHN} \\
 \shortstack{Bilateral\\Deficit}  &  \shortstack{Aggregate\\Deficits}  & \shortstack{Free\\Trade} & {$\overline{\tau^{*}}$} & {$\Delta W$(\%)} & {$\overline{\tau^{*}}$} & {$\Delta W$(\%)} & {$\overline{\tau^{*}}$} & {$\Delta W$(\%)} \\ 
\cmidrule(lr){1-3} \cmidrule(lr){4-5} \cmidrule(lr){6-7} \cmidrule(lr){8-9}
 Yes & Yes & No & 10.36 & 0.025 & 10.41 & 0.008 & 17.04 & -0.138 \\ 
 Yes & No & No & 9.30 & 0.017 & 9.40 & 0.011 & 11.85 & -0.072 \\ 
 No & Yes & No & 11.05 & 0.011 & 9.36 & -0.053 & 16.08 & -0.070 \\ 
 No & No & No & 8.75 & 0.015 & 8.99 & -0.019 & 12.85 & -0.065 \\ 
 \cmidrule(lr){1-3} \cmidrule(lr){4-5} \cmidrule(lr){6-7} \cmidrule(lr){8-9}
 Yes & Yes & Yes & 9.17 & 0.039 & 9.30 & -0.003 & 14.74 & -0.132 \\ 
 Yes & No & Yes & 8.20 & 0.029 & 8.44 & -0.006 & 9.71 & -0.070 \\ 
 No & Yes & Yes & 7.89 & 0.030 & 8.47 & -0.094 & 13.37 & -0.019 \\ 
 No & No & Yes & 7.77 & 0.027 & 8.16 & -0.059 & 10.44 & -0.037 \\ 
\bottomrule
\end{tabular}
\end{center}
\end{table}

Table \ref{tbl:Multisector_Full_results} presents trade-weighted average tariff rates and counterfactual welfare changes for model specifications with and without service sectors and input-output linkages.\footnote{In Appendix \ref{subsec:NumElast} we provide the weighted average numerical elasticities for each of these model specifications, as well as deviations from the weighted Armington elasticity. The results are in line with Section \ref{sec:toy_model}.}  We report trade-weighted average optimal tariff rate and welfare changes relative to both the pre-trade war baseline and a free trade baseline.  Average Nash equilibrium tariff rates range from 9.30--11.83 percent for the United States and 10.71--17.04 percent for China.  Tariff rates and welfare are lowest in the free trade baseline, when services and input-output linkages are included.  This reflects that tariffs incur a spillover cost when placed on goods that serve as intermediate inputs, as they distort production in other sectors.   The United States experiences welfare gains relative to the pre-trade war baseline in all specifications. Excluding services and input-output linkages, however, makes it appear as if the United States would benefit from a trade war even relative to a free trade baseline.  Compared to a framework with no services or input-output linkages, including these features reduces the welfare gains of a trade war from the pre-trade war baseline by 81.8 percent for the United States (from 0.044 percent to 0.008 percent) and increases the welfare losses for China by more than double (133.8 percent; from a loss of 0.059 percent to a loss of 0.138 percent).

\begin{table}[htb]
\caption{Optimal Tariffs and Welfare Changes by Model Structure}
\label{tbl:Multisector_Full_results}
\begin{center}
\begin{tabular}{cccSS|SSSS}
\toprule
\multicolumn{3}{c}{}& \multicolumn{2}{c}{Unilateral Tariffs} & \multicolumn{4}{c}{Nash Tariffs} \\
\cmidrule(lr){4-5} \cmidrule(lr){6-9} 
& &  & \multicolumn{2}{c}{USA} & \multicolumn{2}{c}{USA} & \multicolumn{2}{c}{CHN} \\
IO & Services & \shortstack{Free\\Trade} & {$\overline{\tau^{*}}$} & {$\Delta W$(\%)} & {$\overline{\tau^{*}}$} & {$\Delta W$(\%)} & {$\overline{\tau^{*}}$} & {$\Delta W$(\%)} \\ 
\cmidrule(lr){1-3} \cmidrule(lr){4-5} \cmidrule(lr){6-7} \cmidrule(lr){8-9}
 No & No & No & 9.95 & 0.050 & 10.04 & 0.044 & 10.71 & -0.059 \\ 
 No & Yes & No & 9.32 & 0.012 & 9.66 & 0.010 & 10.79 & -0.033 \\ 
 Yes & No & No & 10.34 & 0.088 & 10.44 & 0.072 & 11.54 & -0.155 \\ 
 Yes & Yes & No & 10.36 & 0.025 & 10.41 & 0.008 & 17.04 & -0.138 \\ 
\cmidrule(lr){1-3} \cmidrule(lr){4-5} \cmidrule(lr){6-7} \cmidrule(lr){8-9}
 No & No & Yes & 11.50 & 0.131 & 11.83 & 0.021 & 13.97 & -0.107 \\ 
 No & Yes & Yes & 9.56 & 0.024 & 9.75 & 0.001 & 13.45 & -0.051 \\ 
 Yes & No & Yes & 11.26 & 0.202 & 11.61 & 0.004 & 15.66 & -0.247 \\ 
 Yes & Yes & Yes & 9.17 & 0.039 & 9.30 & -0.003 & 14.74 & -0.132 \\ 
\bottomrule
\end{tabular}
\end{center}
\end{table}

As expected, across all specifications, welfare is lower for the United States and tariffs are higher in the Nash equilibrium compared to unilateral tariffs --- highlighting the offsetting impact of retaliation inherent in trade wars.  Notably, except in the pre-trade baseline when services are excluded, a tariff war results in China setting higher tariffs on the United States than vice-versa.  This reflects the scenario highlighted by Figure \ref{Fig--Tariff_Rates_3Country_AggImb}, in which each country has imbalances with other countries.  This result is also in part due to the United States' trade surplus in service sectors, and the inability of China to place tariffs on these sectors.  It is possible that allowing for non-tariff trade measures might counteract these effects.  Also notably, across all specifications China loses a trade war with the United States; whereas, the United States only loses a trade war relative to a hypothetical free trade baseline, and only after accounting for service sectors and input-output linkages.

\subsection{Welfare Effects of 2018 Tariff War}
\label{subsec:policy_eval}
Our results suggest that the United States would gain from a trade war with China involving a move towards Nash Equilibrium tariffs.  A natural follow-up question is whether the United States did, in fact, experience welfare gains as a result of the 2018 trade war.  If tariffs were set non-optimally, the resulting quantitative impacts might differ substantially from those discussed in the previous section, in which tariffs were chosen to maximize welfare.  

We evaluate the welfare impact of the trade war between the United States and China by calibrating the model to the 2014 baseline and then solving the counterfactual model with 2019 tariffs.  In the TRAINS tariff data, many sectors experienced small tariff decreases between 2014 and 2019 as part of a longer term trend of lowering tariffs.  To isolate the welfare impact of the trade war escalation in tariffs, our counterfactual includes only tariff increases between 2014 and 2019, leaving tariffs unchanged in other sectors. Table \ref{tbl:Trump_Tariffs_results} reports the welfare changes that resulted from tariff increases that occurred during the trade war.  The results under "US Tariffs Only" report the welfare changes that result from U.S. tariff increases on imports from China, under the assumption that Chinese tariffs on U.S. exports remain at their pre-trade war levels.  The results under "US \& CHN Tariffs" report the welfare changes after  accounting for Chinese tariff increases on imports from the United States.  The results show that the United States experienced welfare losses across all model specifications, even prior to accounting for retaliation from China.  Chinese tariff increases, in contrast, had negligible impacts on both U.S. and Chinese welfare, but did not lead to further welfare losses for China. 

\begin{table}[htb]
\caption{Welfare Changes ($\Delta W$\%) from US-China Trade War, \\ by Model Structure (2019 Tariff Rate Increases) }
\label{tbl:Trump_Tariffs_results}
\begin{center}
\begin{tabular}{ccSSSS}
\toprule
& & \multicolumn{2}{c}{US Tariffs Only} & \multicolumn{2}{c}{US \& CHN Tariffs} \\
\cmidrule(lr){3-4} \cmidrule(lr){5-6}
{IO} & {Services} & {USA} & {CHN} & {USA} & {CHN} \\
\midrule
 No & No & -0.008 & -0.003 & -0.009 & -0.003 \\ 
 No & Yes & -0.002 & -0.002 & -0.002 & -0.002 \\ 
 Yes & No & -0.013 & -0.009 & -0.014 & -0.009 \\ 
 Yes & Yes & -0.003 & -0.009 & -0.004 & -0.008 \\ 
\bottomrule
\end{tabular}
\end{center}
\end{table}

These results, particularly when contrasted with the results in Table \ref{tbl:Optimal_Taus_Full}, suggest that neither country set tariffs to maximize welfare.   The correlation between the United States tariff rate increases during the trade war and the changes implied by Nash equilibrium tariffs reported in Table \ref{tbl:Optimal_Taus_Full} are negative and equal to $-0.09$.  The correlation for China's tariff rate increases is somewhat higher at $0.29$.  If tariff rate increases during the trade war were set primarily for political reasons, this suggests that the United States was willing to incur economic losses for political gains, whereas China's response was closer to economically neutral.  Our results on economic neutrality are consistent with those from \citet{Fetzer2021} who find that China's retaliatory tariffs were politically targeted, rather than set to mitigate the welfare losses from the United States' tariffs.  In contrast, they find that the European Union placed a larger emphasis on mitigating welfare losses from U.S. tariff hikes; proxied for by whether retaliatory tariffs were placed on products for which the United States had a revealed comparative advantage.  These results are also consistent with \citet{KimMargalit2021}, who argue China's response aimed to maximize electoral damage to Trump.

\begin{figure}[htp]
\caption{Welfare Implications of a Unilateral Tariff Rate Reduction by China \label{Fig--CHN_UnilateralTariff_Reduction} } %
	\begin{minipage}[b]{\linewidth}
		\centering
		 	\includegraphics[scale=0.9]{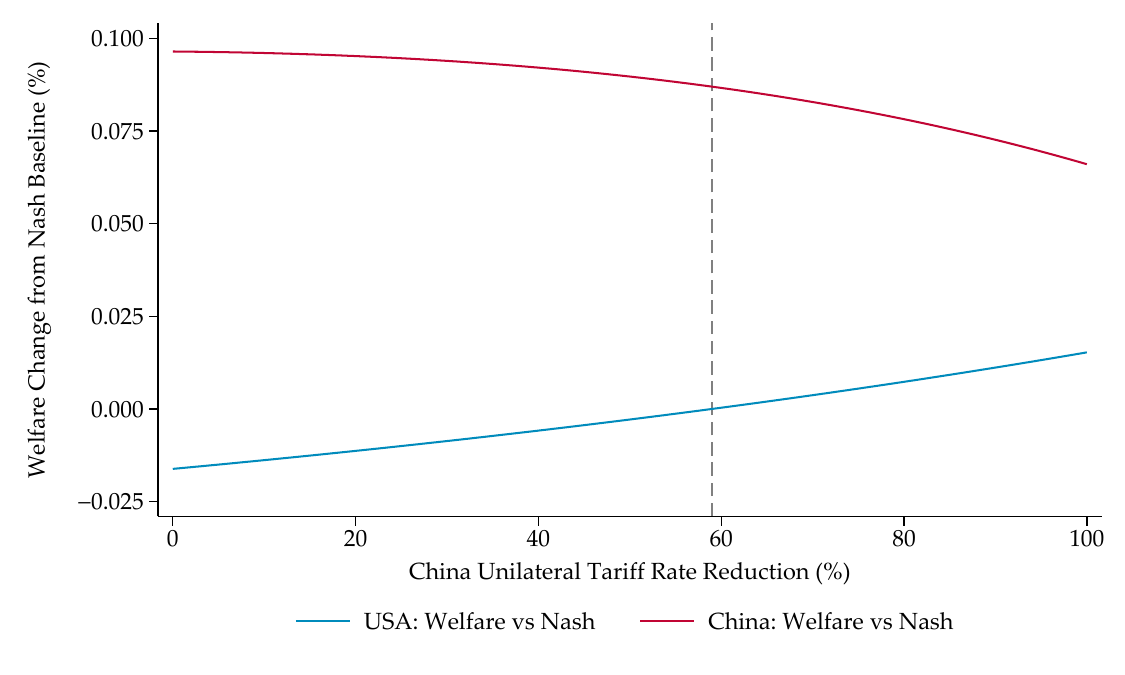}
	\end{minipage}

\end{figure}

Given that the United States benefits from a trade war compared to the baseline in the data, and that both the United States and China are better off under a free trade regime compared to a trade war, one more follow up question is whether China can act unilaterally to avert a trade war.  We start from the pre-trade war baseline in 2014 and compute the counterfactual equilibrium for a uniform, multiplicative, tariff rate reduction applied to all tariffs.  Figure \ref{Fig--CHN_UnilateralTariff_Reduction} displays the consumption equivalent welfare computed using equation \eqref{eq:cons_equiv}, where the changes are relative to the Nash Equilibrium tariffs reported in Table \ref{tbl:Optimal_Taus_Full}.  We find that China would need to unilaterally reduce their tariffs on the United States by 59 percent to make the United States indifferent between baseline tariffs and the Nash equilibrium tariffs that would result from a trade war.  For example, Table \ref{tbl:data_sector_taus_table} shows that China had a pre-trade war tariff of 13.47 percent imports of Motor Vehicles from the United States.  A 59 percentage reduction in tariffs implies reducing the tariff rate to 5.52 percent --- which is still more than double the tariff rate the United States placed on imports of Motor Vehicles from China.  Notably, China would have 0.087 percent higher welfare under this scenario compared to the Nash equilibrium baseline, meaning such a unilateral tariff reduction would be welfare enhancing if used to avert a trade war.  

\begin{table}[htb]
\caption{Welfare Changes ($\Delta W$\%) from US-China Trade War, \\ Effect of Bilateral and Aggregate Imbalances (2019 Tariff Rate Increases) }
\label{tbl:Trump_Tariffs_scenario_results}
\begin{center}
\begin{tabular}{ccSSSS}
\toprule
& & \multicolumn{2}{c}{US Tariffs Only} & \multicolumn{2}{c}{US \& CHN Tariffs} \\
\cmidrule(lr){3-4} \cmidrule(lr){5-6}
{US-CHN Deficit} & {Aggregate Deficits} & {USA} & {CHN} & {USA} & {CHN} \\
\midrule
 Yes & Yes & -0.003 & -0.009 & -0.004 & -0.008 \\ 
 No & Yes & -0.004 & -0.006 & -0.009 & 0.001 \\ 
 Yes & No & -0.003 & -0.005 & -0.004 & -0.005 \\ 
 No & No & -0.004 & -0.006 & -0.009 & 0.000 \\ 
\bottomrule
\end{tabular}
\end{center}
\end{table}

How did aggregate and bilateral imbalances alter the welfare implications of the trade war?  Table \ref{tbl:Trump_Tariffs_scenario_results} reports the welfare changes that would have occurred with the same tariff changes, but starting from counterfactual baselines without bilateral or aggregate trade imbalances.   As before, we eliminate bilateral imbalances between the United States and China by scaling $\gamma_{CHN,USA}^{s}$ proportionally in all sectors (we rebalance other taste shares so we have $\sum_{j=1}^{J|}{\gamma_{CHN,j}^{s}}=1$) until the resulting counterfactual equilibrium has no bilateral deficit between the two countries.  We likewise eliminate aggregate trade imbalances by setting $D_{i}=0$ for all countries. Our results show that bilateral imbalances between the United States and China drive our results much more than aggregate trade imbalances.  The bilateral trade imbalance reduces the welfare loss of the United States by 56 percent ($-0.009$ to $-0.004$) and changes what would be a small welfare gain for China into a loss that is twice that of the United States' loss.

\subsection{Trade Wars with Other Trading Partners}
\label{subsec:partners}
In this section we explore how the United States and China would fare in bilateral trade wars with other countries. Table \ref{tbl:Partner_USA_results} presents the optimal unilateral and Nash equilibrium tariffs, and resulting welfare changes, for a trade war between the United States and each of its trading partners.  The results are for an aggregated trade model featuring three sectors, Agriculture, Commodities, and Manufacturing; and with input-output linkages across sectors.\footnote{We group sectors 1--3 in Agriculture, sectors 4 and 10 in Commodities, and sectors 5--9 and 12--22 in Manufacturing (see Table \ref{tbl:data_sector_taus_table}).  For specifications with services, we aggregate all service sectors together into a single sector.}  This aggregation is meant to reduce the computational time necessary to run the exercise, and we ensure through calibration our baseline remains consistent.  We discuss model aggregation further in Appendix \ref{subsec:aggregation}.  We also exclude services and input-output linkages, which slow convergence to the Nash equilibrium tariffs.  %

\begin{table}[htb]
\caption{Optimal Tariffs and Welfare Changes by Trading Partner, USA}
\label{tbl:Partner_USA_results}
\begin{center}
\begin{tabular}{cSSSSSS}
\toprule
 & \multicolumn{2}{c}{Unilateral Tariffs} & \multicolumn{4}{c}{Nash Tariffs} \\
\cmidrule(lr){2-3} \cmidrule(lr){4-7}
   & \multicolumn{2}{c}{USA} & \multicolumn{2}{c}{USA} & \multicolumn{2}{c}{Partner} \\
\cmidrule(lr){2-3} \cmidrule(lr){4-5} \cmidrule(lr){6-7}
 Partner & {$\overline{\tau^{*}}$} & {$\Delta W$(\%)} & {$\overline{\tau^{*}}$} & {$\Delta W$(\%)} & {$\overline{\tau^{*}}$} & {$\Delta W$(\%)} \\ 
\cmidrule(lr){1-1} \cmidrule(lr){2-3} \cmidrule(lr){4-5} \cmidrule(lr){6-7}
 AUS & 10.63 & 0.003 & 10.77 & -0.013 & 6.51 & -0.022 \\ 
 BRA & 10.59 & 0.013 & 10.79 & -0.002 & 12.55 & -0.157 \\ 
 CAN & 15.79 & 0.296 & 16.74 & 0.028 & 9.55 & -3.928 \\ 
 CHE & 11.31 & 0.006 & 11.31 & 0.004 & 7.94 & -0.443 \\ 
 CHN & 12.88 & 0.148 & 13.09 & 0.048 & 18.13 & -0.307 \\ 
 EUC & 12.27 & 0.176 & 12.75 & -0.077 & 14.38 & -0.319 \\ 
 GBR & 10.95 & 0.015 & 10.97 & 0.013 & 3.85 & -0.183 \\ 
 IDN & 11.32 & 0.009 & 11.32 & 0.006 & 10.62 & -0.180 \\ 
 IND & 11.38 & 0.017 & 11.38 & 0.017 & 11.42 & -0.153 \\ 
 JPN & 12.05 & 0.046 & 12.21 & 0.003 & 8.92 & -0.303 \\ 
 KOR & 11.59 & 0.036 & 11.82 & -0.006 & 14.04 & -0.625 \\ 
 MEX & 15.13 & 0.208 & 16.61 & -0.038 & 12.20 & -3.428 \\ 
 NOR & 10.55 & 0.002 & 10.61 & -0.000 & 7.64 & -0.151 \\ 
 RUS & 10.88 & 0.008 & 10.88 & 0.008 & 7.88 & -0.092 \\ 
 TUR & 10.87 & 0.004 & 10.91 & 0.000 & 9.00 & -0.131 \\ 
 TWN & 11.62 & 0.011 & 11.74 & -0.002 & 11.29 & -0.728 \\ 
\bottomrule
\end{tabular}
\end{center}
\end{table}

Our results show that the United States gains the most from a trade war with China, followed by Canada and India.  The United States loses the most from a trade war with the European Union and with Mexico, even though it would gain significantly from being able to impose unilateral tariffs on each of these trading partners were there no threat of retaliation.  There are no countries for which a trade war with the United States would leave that country better off.  These results may help explain the United States' recent trade policy, characterized by a series of trade conflicts with a variety of trading partners.

\begin{table}[htb]
\caption{Optimal Tariffs and Welfare Changes by Trading Partner, China}
\label{tbl:Partner_CHN_results}
\begin{center}
\begin{tabular}{cSSSSSS}
\toprule
 & \multicolumn{2}{c}{Unilateral Tariffs} & \multicolumn{4}{c}{Nash Tariffs} \\
\cmidrule(lr){2-3} \cmidrule(lr){4-7}
   & \multicolumn{2}{c}{CHN} & \multicolumn{2}{c}{CHN} & \multicolumn{2}{c}{Partner} \\
 \cmidrule(lr){2-3} \cmidrule(lr){4-5} \cmidrule(lr){6-7}
 Partner & {$\overline{\tau^{*}}$} & {$\Delta W$(\%)} & {$\overline{\tau^{*}}$} & {$\Delta W$(\%)} & {$\overline{\tau^{*}}$} & {$\Delta W$(\%)} \\ 
\cmidrule(lr){1-1} \cmidrule(lr){2-3} \cmidrule(lr){4-5} \cmidrule(lr){6-7}
 AUS & 13.66 & 0.026 & 13.80 & 0.009 & 6.99 & -0.822 \\ 
 BRA & 13.97 & 0.005 & 13.94 & 0.006 & 13.21 & -0.140 \\ 
 CAN & 14.71 & 0.003 & 14.85 & -0.017 & 6.15 & -0.063 \\ 
 CHE & 13.26 & 0.001 & 13.26 & -0.002 & 7.83 & -0.152 \\ 
 EUC & 13.64 & 0.102 & 13.99 & -0.130 & 11.33 & -0.117 \\ 
 GBR & 12.97 & 0.001 & 13.11 & -0.017 & 6.03 & -0.027 \\ 
 IDN & 12.69 & 0.009 & 12.59 & -0.015 & 10.80 & -0.098 \\ 
 IND & 12.50 & 0.001 & 12.62 & -0.009 & 13.05 & -0.026 \\ 
 JPN & 13.68 & 0.009 & 14.21 & -0.122 & 11.55 & -0.023 \\ 
 KOR & 15.14 & 0.025 & 15.60 & -0.028 & 13.30 & -1.033 \\ 
 MEX & 13.72 & 0.002 & 13.78 & -0.008 & 8.37 & -0.054 \\ 
 NOR & 13.32 & 0.001 & 13.35 & -0.001 & 6.17 & -0.075 \\ 
 RUS & 10.95 & 0.006 & 11.83 & -0.019 & 8.62 & -0.095 \\ 
 TUR & 11.54 & 0.001 & 11.69 & -0.011 & 8.87 & 0.024 \\ 
 TWN & 17.60 & 0.032 & 17.87 & 0.003 & 11.86 & -3.082 \\ 
 USA & 17.34 & 0.019 & 18.10 & -0.306 & 13.09 & 0.048 \\ 
\bottomrule
\end{tabular}
\end{center}
\end{table}

Table \ref{tbl:Partner_CHN_results} shows the results of the same exercise for trade wars involving China.  In contrast to the United States, China loses from a trade war with most of its trading partners; gaining only from trade wars with Australia, Brazil, and Taiwan.  By far the most damaging trade war for China is with the United States, followed by the European Union, and then Japan.  These results suggest that China is unlikely to be an instigator of trade wars in the near-term, as it stands to lose from most such conflicts.

\section{Concluding Remarks}
\label{conclusion}
In this paper, we highlight the interaction between trade imbalances and international trade policy.  While trade imbalances are often presumed to arise due to domestic macroeconomic policy --- a presumption we do not challenge --- we show that trade imbalances have important implications for the welfare impact of tariffs and for optimal tariff policy.  

Our findings, derived from a multi-sector, multi-region applied general equilibrium model, reveal that countries with significant trade deficits are better positioned to gain from trade wars. Quantitatively, these gains prove to be minimal when applied to a trade war between the two countries with the largest bilateral trade imbalance, the United States and China.  Our analysis underscores the importance of including input-output linkages and service sectors in models of trade policy, as these features reduce the potential gains of a trade war for the United States. 

Our paper demonstrates the utility of global optimization heuristics for solving computationally complex problems.  Economists are likely to find these tools useful for a range of problems, particularly in contexts where solutions are difficult to find, but less costly to verify.  Future work might adapt our methodology for computing optimal unilateral and Nash equilibrium tariffs to study alternative trade policy scenarios.  For example, the optimization criteria in our algorithm could be modified to consider scenarios in which a country seeks to maximize economic damage dealt to a trading partner while minimizing domestic costs.  This might be done, for instance, with the goal of deterring potential instigators of future trade wars.  In this vein, our framework could be extended to examine how trade imbalances affect the formation and stability of multilateral trade agreements. 

Our findings suggest that increasing global imbalances may make it more difficult to maintain international cooperation in trade policy and undermine multinational efforts to decrease tariffs. Large deficits make repudiation of tariff agreements more economically attractive to member states.  Alternatively, trade deficits may present a credible threat of tariff escalations, and thereby encourage tariff reductions to avoid future trade wars.  Regardless, it seems prudent for policymakers to consider the impact of global imbalances, and factors that might exacerbate them, when designing and negotiating multilateral agreements.

\clearpage

\linespread{1.0}
\bibliographystyle{style1}
\bibliography{bib_master}

\newpage
\appendix
\section{Appendix: Ricardian Model}
\label{appendix:CP}
\counterwithin{table}{section}
\counterwithin{figure}{section}
\counterwithin{algorithm}{section}
\setcounter{algorithm}{0}

\label{subsec:cp}
In this section, we show our results are similar when using the multi-sector, multi-country Ricardian framework of \citeauthor{caliendoEstimatesTradeWelfare2015} (\citeyear{caliendoEstimatesTradeWelfare2015}, henceforth ``the CP model").  The CP model is a multi-sector extension of the Eaton-Kortum \citep{eatonTechnologyGeographyTrade2002} trade model and uses the exact-hat notation from \citet{Dekle2008}.

\subsection{Model Differences}
We closely follow the structure and model of the CP model, therefore our focus here is on how this framework differs, and is similar to, our baseline Armington model in Section \ref{sec:model} (henceforth ``the baseline model").

\subsubsection*{Households}
The baseline model and the CP model have the same household structure, where preferences are given by equation (\ref{Utility}) and the budget constraint and income are given by equations (\ref{BC}) and (\ref{BC2}).

\subsubsection*{Differentiated Output}
In the CP model, the Final Goods Bundler produces sectoral output according to the \cite{dixitMonopolisticCompetitionOptimum1977} aggregator over differentiated varieties.  All countries can produce each variety, however, they may face different costs in doing so.  For a particular variety, the output of each country will be perfectly substitutable, which ensures consumers will only consume output from the the lowest cost producer of that variety, 
\begin{equation}
\label{YdefCP}
y_{i}^{s}= \left(  \int \left( r_{i}^{s}(\omega^s) \right) ^{\frac{ \sigma _{s}-1}{ \sigma _{s}}} d \omega^s\right) ^{\frac{ \sigma _{s}}{ \sigma _{s}-1}},    
\end{equation} 
where $r_{i}^{s}(\omega^s)$ is the demand of variety $(\omega^s)$ from the lowest cost supplier. The solution to the problem is 
\begin{equation}
r_{i}^{s}(\omega^s)=\left(\frac{p_{i}^{s}(\omega^s)}{p_{i}^{s}}\right)^{\sigma_s}y_{i}^{s},
\end{equation}
where 
\begin{equation}
\label{pidefCP}
p_{i}^{s} \equiv \left(\int  (p_{ij}^{s}(\omega^s))^{1-\sigma_s}d\omega^s\right) ^ \frac{1}{1-\sigma_s}.
\end{equation}
Equations (\ref{YdefCP}) and (\ref{pidefCP}) are the CP model counterparts to equations (\ref{eq:yij_def}) and (\ref{eq:pi_def}) in the baseline model.

In the CP model, there is a continuum of differentiated varieties that are produced with the following Cobb-Douglas production function,
\begin{equation}
   y_{i}^{s}(\omega^s)= z_{i}^{s}(\omega^s)\left[ l_{i}^{s}(\omega^s)\right]^{\gamma_{i}^{s}}\prod_{k=1}^J\left[m_{ki}^{s}(\omega^s)\right]^{\gamma_{i}^{k,s}},
\end{equation}
where $l_{i}^{s}(\omega^s)$ is labor, $m_{k,i}^{s}$ are the composite intermediate goods from sector $k$ used to produce good $\omega^s$.  Parameters $\gamma_{i}^{k,s}\geq 0$ are the shares of materials used in the production of the good, and $\sum_{k=1}^J \gamma_{i}^{k,s} =1-\gamma_{i}^{s}.$ This is the counterpart to baseline model's equation (\ref{eq:prd_fcn}).  The key difference is the Cobb-Douglas structure in the CP model ensures that intermediate inputs will be used in production according to fixed expenditure shares, rather than in fixed proportions, as in the baseline model.  The unit cost price of variety $\omega^s$ in country $i$ is given by
\begin{equation}
\label{cis}
c_{i}^{s}=\left(\frac{w_i}{\gamma_{i}^{s}}\right)^{\gamma_{i}^{s}}\left[\prod_{k=1}^J\left(\frac{p_i^k}{\gamma_{i}^{k,s}}\right)^{\gamma_{i}^{k,s}}\right].
\end{equation}

There are ad valorem tariffs \(  \tau_{ij}^{s} - 1 \) and iceberg trade costs $t_{ij}^s$, which make the cost of shipping goods between $i$ and $j$ equal to $\kappa_{ij}^s=\tau_{ij}^{s}t_{ij}^s$. This implies that different countries will have different lowest cost producers.  The cost of variety $\omega^s$ in country $i$ is therefore 
\begin{equation}
p_{i}^{s}(\omega^s)=\min_j\left\{\frac{c_j^s\kappa_{ij}^s}{z_j^s(\omega^s)}\right\}.
\end{equation}

\subsubsection*{Trade Shares}
The CP model is based on the multi-product Ricardian framework of \cite{eatonTechnologyGeographyTrade2002}.  We assume that the productivity for producing $\omega^s$ in each country is a realization of a Fr\'{e}chet distribution with location parameter $\lambda_{i}^{s}$ and shape parameter $\theta^s$. This distribution is max-stable, which implies that the maximum productivity, and therefore minimum cost, will also follow a Fr\'{e}chet distribution.  This allows us to focus on the proportion of varieties that will be imported from each trading partner, and allows us to construct price indices for the bundled sectoral output.

In particular, this structure implies that the price index in equation (\ref{pidefCP}) becomes,
\begin{equation}
\label{pidefCP2}
p_{i}^{s} = A^s \left(\sum_{n=1}^{N} \lambda_{n}^{s} \left(c_{n}^{s}\kappa_{in}^{s}\right)^{-\theta^s}\right) ^ \frac{-1}{\theta^s},
\end{equation}
where $A^s$ is a constant, and the final good purchased by households becomes $P_i=\prod_{s=1}^S \left(P^s_i / \alpha^s_i\right)^{\alpha^s_i}.$ Equation (\ref{pidefCP2}) is the CP model counterpart to equation (\ref{eq:pij_sol}) in the baseline model.

We define the expenditure in sector $s$ goods in country $i$ as $X_i^s=P_i^sQ_i^s$, and the expenditure share, $\pi_{in}^{s}$, of country $i$'s, sector $s$ from country $n$ as
\begin{equation}
\pi_{in}^{s}=\frac{\lambda_{n}^{s} \left(c_{n}^{s}\kappa_{in}^{s}\right)^{-\theta^s}}{\sum_{n=1}^{N} \lambda_{n}^{s} \left(c_{n}^{s}\kappa_{in}^{s}\right)^{-\theta^s}},
\end{equation}
and the expenditure on goods as
\begin{equation}
X_{i}^{s}=\sum_{k=1}^J\gamma_i^{s,k}\sum_{n=1}^NX_n^k\frac{\pi_{ni}^{k}}{\tau_{ni}^{k}}+\alpha_i^s{I}_i.
\end{equation}
These expressions allow us to write the government budget constraint as
\begin{equation}
T_i=\sum_{s=1}^S\sum_{j=1}^J(\tau_{ij}^s-1)\frac{\pi_{in}^{s}X_i^s}{\tau_{ij}^s},
\label{eq:tariffrevenueCP}
\end{equation}
where equation (\ref{eq:tariffrevenueCP}) is the CP model counterpart of equation (\ref{eq:tariffrevenue}), and to write the aggregate trade imbalance as 
\begin{equation}
D_i=\sum_{s=1}^S\left(
\sum_{j=1}^J\frac{\pi_{in}^{s}X_i^s}{\tau_{ij}^s}
-
\sum_{j=1}^J\frac{\pi_{jn}^{s}X_i^s}{\tau_{ji}^s}\right),
\label{eq:tradebalanceCP}
\end{equation}
where equation (\ref{eq:tradebalanceCP}) is the CP model counterpart of equation (\ref{eq:tradebalance}).

Finally, the goods market clear, so that total output equals final demand plus intermediate input usage,
\begin{equation}
y_i^s=c_i^s+\sum _{n=1}^{N} \int m_{ni}^{s}(\omega^s) d\omega^s.
\label{eq:goodmarketCP}
\end{equation}
Here, equation (\ref{eq:goodmarketCP}) is the counterpart to equation (\ref{eq:goodmarket}).  

We follow the same calibration procedure as in \citet{caliendoEstimatesTradeWelfare2015}, which is largely the same as in Section \ref{sec:calibration}.  Tariff revenues, trade deficits, income, and final consumption share parameters are calibrated with the same exact equations and data as in the baseline model.  The only remaining objects to calibrate are the trade shares, $\pi_{in}^{s}$, which are calibrated to match expenditure shares in the data (expenditure flow divided by expenditures summed across sources, inclusive of tariffs), and the input-output parameters.  The input-output parameters are calibrated to be equal to the IO flow in the data divided by expenditures summed across destinations, net of tariffs.  Note that these input-output parameters differ only due to the Cobb-Douglas specification; otherwise, they would be the same as in the Armington framework as well.

\subsubsection*{Equilibrium in Changes}
We implement the exact-hat notation from \citet{Dekle2008}, similar to section 3.1.7 in \citet{caliendoEstimatesTradeWelfare2015}, where we label as $\hat{x}$ the relative change of variable $x$ between two policies, $\tau'$ (characterized by equilibrium prices $(w',p')$) and $\tau$ (characterized by equilibrium prices $(w,p)$) --- namely, $\hat{x}=x'/x$. The equilibrium in relative changes is given by equations:

\noindent
Cost of input bundles:
\begin{equation}
\hat{c}_{i}^{s}=\left(\frac{\hat{w}_i}{\gamma_{i}^{s}}\right)^{\gamma_{i}^{s}}\left[\prod_{k=1}^J\left(\frac{\hat{p}_i^k}{\gamma_{i}^{k,s}}\right)^{\gamma_{i}^{k,s}}\right].
\end{equation}
Price Index:
\begin{equation}
\hat{p}_{i}^{s} = \left(\sum_{n=1}^{N} \pi_{in}^{s} \left(\hat{c}_{n}^{s}\hat{\kappa}_{in}^{s}\right)^{-\theta^s}\right) ^ \frac{-1}{\theta^s}.
\end{equation}
Bilateral trade shares:
\begin{equation}
\hat{\pi}_{in}^{s}=\left[\frac{\hat{c}_{n}^{s}\hat{\kappa}_{in}^{s}}{\hat{p}_{i}^{s}}\right]^{-\theta^s}.
\end{equation}
Expenditure:
\begin{equation}
X_{i}^{s'}=\sum_{k=1}^J\gamma_i^{s,k}\sum_{n=1}^NX_n^k\frac{\pi_{ni}^{k'}}{\tau_{ni}^{k'}}+\alpha_i^s{I}_i'.
\end{equation}
Trade Balance:
\begin{equation}
D_i=\sum_{s=1}^S\left(
\sum_{j=1}^J\frac{\pi_{in}^{s'}X_i^{s'}}{\tau_{ij}^{s'}}
-
\sum_{j=1}^J\frac{\pi_{jn}^{s'}X_i^{s'}}{\tau_{ji}^{s'}}\right),
\end{equation}
where $\hat{\kappa}_{in}^{s}=\tau_{ij}^{s'}/\tau_{ij}^{s}$ and ${I}_i' = w_{i}'L_{i}+\sum_{s=1}^S\sum_{j=1}^J(\tau_{ij}^{s'}-1)\frac{\pi_{in}^{s'}X_i^{s'}}{\tau_{ij}^{s'}}+D_{i}$.

After solving for the equilibrium in changes, it is possible to calculate the change in welfare as
\begin{equation}
   \hat{W}_i = \frac{\hat{I}_i}{\prod_{s=1}^{S} \left( \hat{p}_{i}^{s} \right)^{a_{i}^{s}}}.
\end{equation}
Following calibration, we use the same algorithm as in Section \ref{sec:optimal_tariffs_algorithm} to find the optimal unilateral tariffs and Nash equilibrium tariffs in the CP model.  Note that ${\text{W}}_i' = \hat{\text{W}}_i \text{W}_i$, therefore, finding the tariffs that maximize welfare is equivalent to finding the tariffs that maximize the change in welfare; and no adjustments are needed to our algorithm when computing optimal tariffs with an equilibrium in changes approach.  

\subsection{Results}
Table \ref{tbl:Optimal_Taus_CP} reports the optimal unilateral and Nash equilibrium tariffs for the CP model.  We report optimal tariffs for both a pre-trade war baseline as well as for a free trade baseline.  This table is the counterpart to Table \ref{tbl:Optimal_Taus_Full} in Section \ref{sec:application}.  %

\begin{table}[!htb]
\caption{Optimal Tariffs Between United States and China\\(CP Model with Services and Input-Output Linkages)}
\label{tbl:Optimal_Taus_CP}
\begin{center}
\begin{tabular}{cSSSSSS}
\toprule
& \multicolumn{3}{c}{Pre-Trade War Baseline} & \multicolumn{3}{c}{Free Trade Baseline} \\
\cmidrule(lr){2-4} \cmidrule(lr){5-7}
 & {Unilateral} & \multicolumn{2}{c}{Nash Tariffs} & {Unilateral} & \multicolumn{2}{c}{Nash Tariffs} \\
  \cmidrule(lr){2-2} \cmidrule(lr){3-4} \cmidrule(lr){5-5} \cmidrule(lr){6-7}
{Sector} & {USA} & {USA} & {CHN} & {USA} & {USA} & {CHN} \\
\midrule
{1} & 8.76 & 9.68 & 12.57 &  9.88 & 10.83 & 12.16 \\ 
{2} & 8.94 & 9.63 & 14.55 &  9.98 & 10.71 & 13.07 \\ 
{3} & 6.01 & 6.77 & 10.23 &  6.08 & 6.99 & 9.67 \\ 
{4} & 5.35 & 6.11 & 8.37 &  5.28 & 6.15 & 7.72 \\ 
{5} & 7.31 & 8.16 & 10.68 &  7.21 & 8.12 & 9.75 \\ 
{6} & 6.05 & 6.78 & 12.97 &  2.77 & 3.52 & 12.40 \\ 
{7} & 6.63 & 7.44 & 10.55 &  6.58 & 7.48 & 9.70 \\ 
{8} & 6.66 & 7.47 & 10.67 &  6.70 & 7.56 & 9.41 \\ 
{9} & 7.62 & 8.46 & 10.84 &  7.49 & 8.32 & 10.03 \\ 
{10} & 6.18 & 6.93 & 9.99 &  6.07 & 6.89 & 9.02 \\ 
{11} & 6.50 & 7.25 & 11.06 &  6.18 & 7.00 & 9.58 \\ 
{12} & 6.04 & 6.86 & 9.35 &  5.98 & 6.89 & 8.83 \\ 
{13} & 6.31 & 7.10 & 11.04 &  6.14 & 6.99 & 9.73 \\ 
{14} & 6.54 & 7.34 & 9.62 &  6.33 & 7.18 & 8.78 \\ 
{15} & 4.81 & 5.57 & 9.50 &  4.74 & 5.53 & 8.28 \\ 
{16} & 6.65 & 7.44 & 10.42 &  6.62 & 7.45 & 9.29 \\ 
{17} & 5.76 & 6.44 & 13.36 &  5.49 & 6.24 & 9.79 \\ 
{18} & 4.48 & 5.25 & 9.87 &  4.20 & 5.04 & 8.19 \\ 
{19} & 4.97 & 5.74 & 9.02 &  4.78 & 5.63 & 7.40 \\ 
{20} & 5.23 & 6.00 & 9.77 &  4.85 & 5.67 & 8.48 \\ 
{21} & 7.23 & 8.01 & 9.66 &  7.18 & 8.03 & 8.55 \\ 
{22} & 5.81 & 6.61 & 10.91 &  5.54 & 6.44 & 6.23 \\ 
 \midrule \rule{0pt}{2.5ex}  {$\overline{\tau^{*}}$} & 5.74 & 6.48 & 10.71 & 5.03 & 5.83 & 9.20 \\ 
\cmidrule(lr){1-1} \cmidrule(lr){2-2} \cmidrule(lr){3-4} \cmidrule(lr){5-5} \cmidrule(lr){6-7}
 \rule{0pt}{2.5ex}  {$\Delta W(\%)$} & 0.007 & 0.005 & -0.050 & 0.015 & -0.008 & -0.057 \\ 
\bottomrule
\end{tabular}
\end{center}
\end{table}

The results in Table \ref{tbl:Optimal_Taus_CP} are quantitatively similar and qualitatively equivalent to our baseline model.  The main difference is that optimal tariff rates are lower in the CP model, with a weighted average tariff of 10.41 percent in our baseline model in Table \ref{tbl:Optimal_Taus_Full} compared to only 6.48 percent here.  Welfare gains are also somewhat lower, equal to 0.005 here compared to 0.008 in our baseline model.  For both specifications, China has higher optimal tariff rates and greater welfare losses from a trade war.  In both our baseline model and the CP model, the United States gains from a trade war relative to existing tariff rates.  Relative to a baseline with free trade, however, both countries exhibit welfare losses from a trade war.

\section{Appendix: Robustness and Extensions}
In these subsections, we explore the quantitative implications of various modifications to our applied general equilibrium framework.

\begin{table}[!htb]
\caption{Optimal Tariffs Between United States and China\\(Model without Services and without Input-Output Linkages)}
\label{tbl:Optimal_Taus_Full_noIO}
\begin{center}
\begin{tabular}{cSSSSSS}
\toprule
& \multicolumn{3}{c}{Pre-Trade War Baseline} & \multicolumn{3}{c}{Free Trade Baseline} \\
\cmidrule(lr){2-4} \cmidrule(lr){5-7}
 & {Unilateral} & \multicolumn{2}{c}{Nash Tariffs} & {Unilateral} & \multicolumn{2}{c}{Nash Tariffs} \\
  \cmidrule(lr){2-2} \cmidrule(lr){3-4} \cmidrule(lr){5-5} \cmidrule(lr){6-7}
{Sector} & {USA} & {USA} & {CHN} & {USA} & {USA} & {CHN} \\
\midrule
{1} & 5.15 & 11.84 & 10.99 &  15.14 & 15.59 & 14.58 \\ 
{2} & 0.23 & 11.01 & 10.68 &  11.72 & 13.91 & 13.87 \\ 
{3} & 2.81 & 10.46 & 11.31 &  12.82 & 12.89 & 14.55 \\ 
{4} & 9.72 & 10.00 & 9.13 &  12.35 & 12.67 & 13.22 \\ 
{5} & 12.15 & 12.60 & 11.18 &  15.41 & 15.93 & 14.72 \\ 
{6} & 10.05 & 10.10 & 11.14 &  8.04 & 8.32 & 14.45 \\ 
{7} & 12.32 & 12.16 & 11.12 &  15.40 & 15.69 & 15.25 \\ 
{8} & 11.56 & 12.06 & 10.92 &  15.09 & 15.32 & 14.55 \\ 
{9} & 3.39 & 12.91 & 11.12 &  15.28 & 16.32 & 15.61 \\ 
{10} & 11.70 & 12.03 & 10.94 &  14.60 & 15.04 & 14.68 \\ 
{11} & 11.10 & 11.22 & 10.66 &  13.51 & 13.81 & 14.13 \\ 
{12} & 10.43 & 10.35 & 10.70 &  12.79 & 13.68 & 14.70 \\ 
{13} & 11.42 & 11.56 & 11.07 &  14.27 & 14.68 & 14.67 \\ 
{14} & 11.70 & 11.95 & 11.24 &  14.63 & 15.00 & 15.03 \\ 
{15} & 10.43 & 10.43 & 10.43 &  12.90 & 13.23 & 14.50 \\ 
{16} & 11.82 & 11.94 & 11.10 &  14.86 & 15.19 & 14.58 \\ 
{17} & 9.12 & 9.16 & 10.00 &  10.95 & 11.27 & 11.85 \\ 
{18} & 8.80 & 8.86 & 10.91 &  10.66 & 10.97 & 14.18 \\ 
{19} & 10.15 & 10.22 & 10.61 &  12.39 & 12.78 & 13.89 \\ 
{20} & 9.75 & 9.79 & 10.91 &  11.88 & 12.17 & 14.47 \\ 
{21} & 11.36 & 11.45 & 10.88 &  14.31 & 14.45 & 14.64 \\ 
{22} & 10.62 & 10.72 & 10.21 &  13.13 & 13.37 & 10.21 \\ 
 \midrule \rule{0pt}{2.5ex}  {$\overline{\tau^{*}}$} & 9.95 & 10.04 & 10.71 & 11.50 & 11.83 & 13.97 \\ 
\cmidrule(lr){1-1} \cmidrule(lr){2-2} \cmidrule(lr){3-4} \cmidrule(lr){5-5} \cmidrule(lr){6-7}
 \rule{0pt}{2.5ex}  {$\Delta W(\%)$} & 0.050 & 0.044 & -0.059 & 0.131 & 0.021 & -0.107 \\ 
\bottomrule
\end{tabular}
\end{center}
\end{table}

\subsection{Uniform Tariff Rates Across Sectors}
\label{subsec:uniform_tariffs}

Our methodology can be adapted to evaluate the implications of tariff rate regimes that restrict tariff rates to be uniform across sectors.  To do so, we equalize tariff rates across sectors by setting $\tau_{USA,CHN}^{s}=\tau_{USA,CHN}$, and apply our prior methodology to compute optimal unilateral and Nash equilibrium tariffs, given this restriction.  Table \ref{tbl:Optimal_Taus_Uniform} reports optimal unilateral and Nash equilibrium tariffs when tariffs are restricted to be uniform for both the United States and China.  The results are similar to those in Table \ref{tbl:Optimal_Taus_Full}.  Furthermore, we reach the same conclusion that the United States gains from a trade war relative to the pre-trade war baseline, but experiences small welfare losses relative to a free-trade baseline.

\begin{table}[htb]
\caption{Optimal Uniform Tariffs Between United States and China\\(Model with Services and Input-Output Linkages)}
\label{tbl:Optimal_Taus_Uniform}
\begin{center}
\begin{tabular}{cSSSSSS}
\toprule
& \multicolumn{3}{c}{Pre-Trade War Baseline} & \multicolumn{3}{c}{Free Trade Baseline} \\
\cmidrule(lr){2-4} \cmidrule(lr){5-7}
 & {Unilateral} & \multicolumn{2}{c}{Nash Tariffs} & {Unilateral} & \multicolumn{2}{c}{Nash Tariffs} \\
  \cmidrule(lr){2-2} \cmidrule(lr){3-4} \cmidrule(lr){5-5} \cmidrule(lr){6-7}
Sector & USA & USA & CHN & USA & USA & CHN \\
\midrule
{$\tau^{*}$} & 10.27 & 10.34 & 16.94 &  9.10 & 9.28 & 14.38 \\ 
\cmidrule(lr){1-1} \cmidrule(lr){2-2} \cmidrule(lr){3-4} \cmidrule(lr){5-5} \cmidrule(lr){6-7}
   {$\Delta W(\%)$} & 0.025 & 0.008 & -0.137 & 0.038 & -0.004 & -0.133 \\ \hline
\bottomrule
\end{tabular}
\end{center}
\end{table}

Table \ref{tbl:Optimal_Taus_Uniform_alt} reports optimal unilateral and Nash equilibrium tariffs for the United States with and without the restriction that China is forced to enact uniform tariffs itself.  These results show that --- for the specification with input-output linkages, but without services --- uniformity achieves a large majority of the potential welfare gains from optimal tariff setting compared to Table \ref{tbl:Multisector_Full_results}, where we allow tariffs to vary across sectors.  Notably, the welfare implications are nearly identical regardless of whether we restrict China to implementing uniform tariffs or not.  Although China does have higher welfare, and the United States has lower welfare, under the Nash equilibrium when China is not restricted to uniform tariffs, the magnitudes are so small that they are not visible on the table.  These results lend credence to political economy and other arguments for uniform tariffs, for example as argued by \citet{Panagariya1993}.

\begin{table}[htb]
\caption{Optimal Uniform Tariffs Between United States and China}
\label{tbl:Optimal_Taus_Uniform_alt}
\begin{center}
\begin{tabular}{cSSSSSS}
\toprule
& \multicolumn{3}{c}{China Uniform} & \multicolumn{3}{c}{China non-Uniform} \\
\cmidrule(lr){2-4} \cmidrule(lr){5-7}
 & {Unilateral} & \multicolumn{2}{c}{Nash Tariffs} & {Unilateral} & \multicolumn{2}{c}{Nash Tariffs} \\
  \cmidrule(lr){2-2} \cmidrule(lr){3-4} \cmidrule(lr){5-5} \cmidrule(lr){6-7}
Sector & USA & USA & CHN & USA & USA & CHN \\
\midrule
{$\tau^{*}$} & 9.10 & 9.28 & 14.38 &  9.10 & 9.31 & 14.77 \\ 
\cmidrule(lr){1-1} \cmidrule(lr){2-2} \cmidrule(lr){3-4} \cmidrule(lr){5-5} \cmidrule(lr){6-7}
   {$\Delta W(\%)$} & 0.038 & -0.004 & -0.133 & 0.038 & -0.004 & -0.133 \\ \hline
\bottomrule
\end{tabular}
\end{center}
\end{table}

\begin{figure}[htp]
\caption{Welfare Consequences of Deviations from Nash Uniform Tariffs \label{Fig--NashVerify_Uniform} } %
	\begin{minipage}[b]{\linewidth}
		\centering
		 	\includegraphics[scale=0.9]{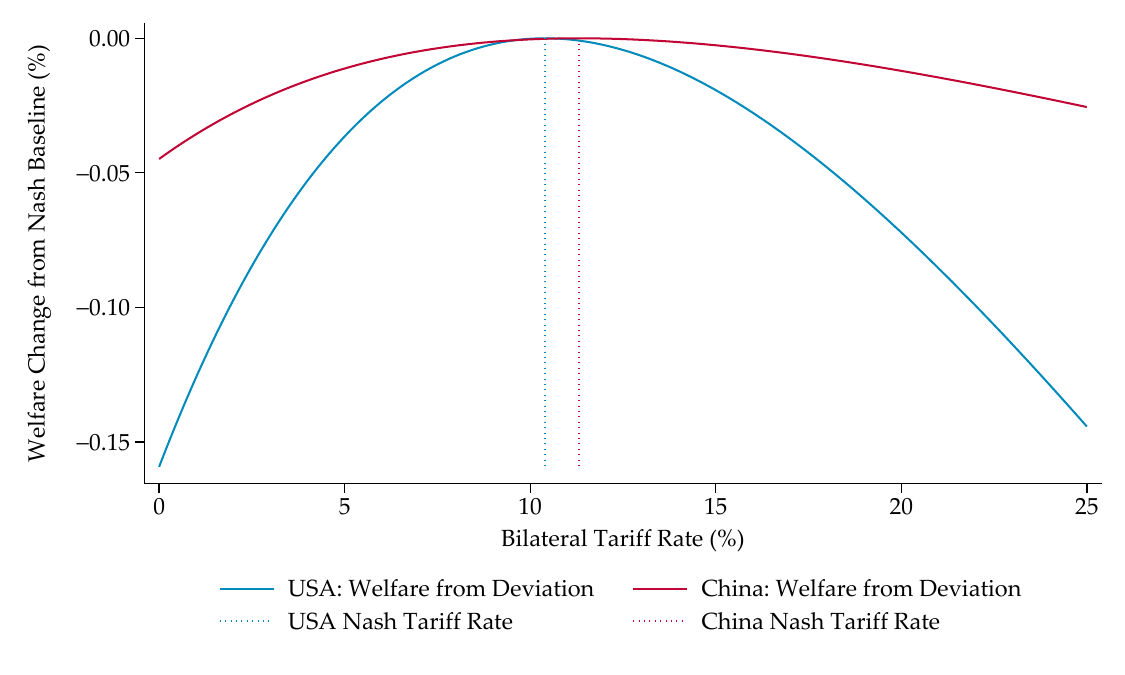}
	\end{minipage}

\end{figure}

It is possible to exhaustively verify that our genetic algorithm converges to a Nash equilibrium when we have uniform tariffs.  We do this by discretizing the grid of uniform tariffs the same as in Section \ref{subsec:ga}, and then evaluating welfare for each point on the grid.  Figure \ref{Fig--NashVerify_Uniform} presents the change in welfare for each tariff rate, assuming that their partner country sets their uniform tariff rate at the Nash equilibrium rate reported in Table \ref{tbl:Optimal_Taus_Uniform_alt}.  Welfare is reported as consumption equivalent variation relative to the baseline in which both countries set tariffs at their Nash equilibrium levels.  Welfare declines as tariff rates further depart from their Nash value, and this pattern continues well beyond the bounds displayed the graph.  This exercise establishes that neither country has an incentive to deviate from the Nash equilibrium, and that our genetic algorithm converges to the correct solution.

\subsubsection{Welfare Impact of Proposed Tariff Hikes}
As part of his 2024 U.S. presidential election campaign, President Trump proposed a uniform tariff rate of 10 to 20 percent on all imports into the United States, with a 60 percent uniform tariff on imports from China.  We evaluate the welfare impact of these tariffs, both if these tariffs were implemented unilaterally and if each country partner respond with equivalent tariffs imposed on the United States.  We also consider two hypothetical situations, in which countries respond to these tariffs by lowering trade barriers with each other, but maintain tariffs at their pre-trade war rates for the United States.  The results for each scenario are reported in Table \ref{tbl:Trump_Unilateral_Tariffs_results}.  

We find that the United States would gain from unilateral tariffs if there is no retaliation.  Across all scenarios, however, the United States would lose if faced with equivalent retaliation.  Equivalent retaliation is unlikely to be optimal, and optimal responses by each partner country would be expected to further exacerbate welfare losses for the United States.  The welfare changes for the Rest of World (ROW) is computed by finding the consumption equivalent applied to mean welfare across countries, which masks a degree of heterogeneity in welfare implications across other countries.  

Notably, we find that increased economic integration between other countries and China would not offset the harm they would face if the proposed tariffs were implemented by the United States, highlighting the importance of the United States' trade policy for the global economy.  

\begin{table}[htb]
\caption{Welfare Changes ($\Delta W$\%) from Proposed Uniform Tariffs \\ by Model Structure (Unilateral Tariffs and with Equal Retaliation) }
\label{tbl:Trump_Unilateral_Tariffs_results}
\begin{center}
\begin{tabular}{ccSSS|SSS}
\toprule
\multicolumn{2}{c}{} & \multicolumn{6}{c}{Pre-Trade War Baseline} \\
\cmidrule[\heavyrulewidth]{3-8}
\multicolumn{2}{c}{} & \multicolumn{3}{c}{US Tariffs Only} & \multicolumn{3}{c}{Equivalent Retaliation} \\
\cmidrule(lr){3-5} \cmidrule(lr){6-8}
IO & Services & {USA} & {CHN} & {ROW} & {USA} & {CHN} & {ROW} \\
\midrule[\heavyrulewidth]
 No & No & 0.987 & -0.196 & -0.477 & -1.023 & -0.211 & -0.309  \\ 
 No & Yes & 0.156 & -0.116 & -0.145 & -0.160 & -0.127 & -0.123  \\ 
 Yes & No & 1.611 & -0.503 & -0.870 & -1.666 & -0.544 & -0.564  \\ 
 Yes & Yes & 0.403 & -0.894 & -0.265 & -0.141 & -0.657 & -0.239  \\ 
\midrule[\heavyrulewidth]
\multicolumn{8}{c}{Tariffs Eliminated Between {All} Countries, Excl. {CHN} and USA} \\
\midrule
 No & No & 0.934 & -0.207 & -0.447 & -1.068 & -0.222 & -0.284  \\ 
 No & Yes & 0.146 & -0.122 & -0.132 & -0.168 & -0.133 & -0.110  \\ 
 Yes & No & 1.526 & -0.531 & -0.822 & -1.740 & -0.573 & -0.523  \\ 
 Yes & Yes & 0.392 & -0.924 & -0.233 & -0.150 & -0.687 & -0.208  \\ 
\midrule[\heavyrulewidth]
\multicolumn{8}{c}{Tariffs Eliminated Between {All} Countries, Incl. CHN, Excl. USA} \\
\midrule
 No & No & 0.915 & -0.218 & -0.385 & -1.078 & -0.236 & -0.223  \\ 
 No & Yes & 0.143 & -0.130 & -0.109 & -0.168 & -0.143 & -0.088  \\ 
 Yes & No & 1.518 & -0.599 & -0.706 & -1.739 & -0.650 & -0.408  \\ 
 Yes & Yes & 0.393 & -0.888 & -0.192 & -0.143 & -0.664 & -0.167  \\ 
\bottomrule
\end{tabular}
\end{center}
\end{table}

\subsection{Bilateral Trade Imbalances from Unobserved Trade Costs}
\label{subsec:iceberg}
In order to rationalize the bilateral trade imbalances we observe in the data, and to be able to calibrate our model to exactly match the data, we implicitly assume that these imbalances arise due to asymmetries in exogenous model parameters.  In Section \ref{sec:calibration} we allow these asymmetries to arise due to differences in preferences across countries through the parameter $\gamma_{i,j}^s$ in equation \eqref{Ydef}.  An alternative presumption is that unobserved trade costs, rather than differences in preferences, might be responsible for trade imbalances.  This is the approach taken by \citet{Cunat2023}.  Without additional data on preferences or non-tariff trade barriers, it is not possible to disentangle the extent to which asymmetries might be caused by one or the other (if we have such data, we can simply include it in equation \eqref{eq:yij_calibration}).  We can, however, explore the robustness of our results to various assumptions on the source of imbalances.  

In the main calibration in the paper, we assume that there are no unobserved iceberg trade costs, i.e. $t_{ij}^{s}=1$ in equation \eqref{eq:prd_fcn}, which allows us to uniquely identify $\gamma_{i,j}^s$. In this section, we instead assume that unobserved iceberg trade costs are responsible for the bilateral trade imbalances, and we calibrate the model to match the data under this assumption.  We then recompute the optimal tariff rates and welfare implications of a trade war between the United States and China.

In order to calibrate iceberg tariffs, we first assume that $t_{jj}^{s} = 1$, meaning there are no additional costs when the source and destination is the same country.  We can then use equation \eqref{eq:yij_calibration} to calibrate $y_{jj}^s$ for each country and sector.  This allows us to infer $p_{jj}^s$ by dividing domestic flows by output, according to equation \eqref{eq:pij_calibration}.  Using equation \eqref{eq:yij_def} for $ij$ and $ii$ and rearranging gives the following expression for prices,
\begin{equation}
    p_{ij}^{s}=p_{ii}^{s} {\left( \frac{p_{ij}^{s}y_{ij}^{s}}{p_{ii}^{s}y_{ii}^{s}}\right)}^{\frac{1}{1-\sigma_{s}}}.
    \label{eq:pij_calibration_iceberg_step1}
\end{equation}
Rearranging equation \eqref{eq:pij_sol}, shows that prices of differentiated output are given by %
\begin{equation} 
     \frac{p_{ij}^{s}}{ \tau_{ij}^{s}t_{ij}^{s}}= \left( \frac{w_{j}}{z_{j}^{s}} \right) + \sum _{k=1}^{S}p_{j}^{k} \alpha _{j}^{sk}. 
     \label{eq:pij_calibration_iceberg_step2}
    \end{equation}
Since $\tau_{jj}^{s}=1$ and $t_{jj}^{s}=1$, this means
\begin{equation} 
    p_{jj}^{s}= \left( \frac{w_{j}}{z_{j}^{s}} \right) + \sum _{k=1}^{S}p_{j}^{k} \alpha _{j}^{sk}. 
   \end{equation}
We can use the above along with equations \eqref{eq:pij_calibration_iceberg_step1} and \eqref{eq:pij_calibration_iceberg_step2} to calibrate $t_{ij}^s$ as
\begin{equation} 
    t_{ij}^{s} = \frac{1}{ \tau_{ij}^{s}} \frac{{ p_{ii}^{s}}}{{ p_{jj}^{s}}}{\left( \frac{p_{ij}^{s}y_{ij}^{s}}{p_{ii}^{s}y_{ii}^{s}}\right)}^{\frac{1}{1-\sigma_{s}}}.
   \end{equation}

   We restrict $t_{ij}^{s} \ge 1$, which ensures it cannot be less costly to deliver one unit of output to a foreign country than to deliver it domestically.  We further restrict $t_{ij}^s\le 100$ to place an upper bound on iceberg trade costs.  These restrictions are meant to prevent implausible trade costs, however, they can be relaxed if desired.  These restrictions mean, for instance, that the model will interpret zeros (origin-destination-sector tuples with zero trade) as arising due to differences in preferences, rather than unobserved trade costs making it infinitely expensive to deliver output in a given sector to a specific destination.  Eliminating these restrictions would force iceberg costs to fully account for global imbalances.  Following the calibration of iceberg trade costs, we are able to plug them into equation \eqref{eq:yij_calibration} and proceed through the remaining steps of the calibration in Section \ref{sec:calibration}.

   \begin{table}[!htb]
\caption{Optimal Tariffs Between United States and China\\(Model with Iceberg Costs, Services, and Input-Output Linkages)}
\label{tbl:Optimal_Taus_Full_iceberg}
\begin{center}
\begin{tabular}{lSSSSSS}
\toprule
& \multicolumn{2}{c}{Pre-Trade War Baseline} & \multicolumn{2}{c}{Free Trade Baseline} & \multicolumn{2}{c}{Frictionless Baseline} \\
\cmidrule(lr){2-3}  \cmidrule(lr){4-5} \cmidrule(lr){6-7}
{Sector} & {USA} & {CHN} & {USA} & {CHN} & {USA} & {CHN} \\
\midrule
1 & 15.34 & 14.14 & 16.13 &  18.53 & 17.98 & 14.37 \\ 
2 & 16.54 & 14.28 & 14.63 &  16.52 & 33.91 & 32.36 \\ 
3 & 10.08 & 12.34 & 10.55 &  16.65 & 6.91 & 5.70 \\ 
4 & 8.86 & 7.81 & 9.43 &  12.20 & 6.40 & 5.75 \\ 
5 & 12.40 & 12.25 & 13.32 &  15.98 & 4.54 & 4.86 \\ 
6 & 10.21 & 14.32 & 6.37 &  18.53 & 3.44 & 6.94 \\ 
7 & 11.98 & 11.40 & 12.74 &  15.22 & 6.78 & 5.95 \\ 
8 & 11.42 & 11.35 & 12.19 &  14.45 & 4.73 & 5.87 \\ 
9 & 12.63 & 11.33 & 13.20 &  15.07 & 1.29 & 2.19 \\ 
10 & 9.63 & 9.19 & 10.10 &  12.56 & 1.88 & 2.69 \\ 
11 & 10.75 & 11.34 & 10.86 &  14.43 & 9.11 & 8.93 \\ 
12 & 10.47 & 10.86 & 10.79 &  15.32 & 5.96 & 4.42 \\ 
13 & 11.17 & 12.07 & 11.67 &  15.19 & 4.98 & 7.63 \\ 
14 & 10.82 & 10.98 & 11.42 &  14.95 & 7.68 & 4.18 \\ 
15 & 9.33 & 10.08 & 9.33 &  13.22 & 5.67 & 5.61 \\ 
16 & 11.49 & 11.16 & 11.79 &  14.54 & 7.94 & 5.03 \\ 
17 & 10.86 & 14.78 & 10.79 &  16.32 & 9.07 & 13.28 \\ 
18 & 8.34 & 11.02 & 8.76 &  14.09 & 11.92 & 5.06 \\ 
19 & 9.16 & 10.07 & 10.09 &  12.94 & 5.32 & 3.76 \\ 
20 & 10.97 & 11.21 & 11.37 &  14.62 & 4.57 & 5.81 \\ 
21 & 12.64 & 10.89 & 13.35 &  14.68 & 4.70 & 4.30 \\ 
22 & 10.33 & 10.45 & 10.99 &  7.84 & 5.92 & 7.47 \\ 
\midrule  {$\overline{\tau^{*}}$} & 10.37 & 11.74 & 10.01 & 14.85 & 7.20 & 7.60 \\ 
\cmidrule(lr){1-1} \cmidrule(lr){2-3}  \cmidrule(lr){4-5} \cmidrule(lr){6-7}
 {$\Delta W(\%)$} & 0.024 & -0.117 & 0.004 & -0.152 & -0.304 & -0.278 \\ \hline
\bottomrule
\end{tabular}
\end{center}
\end{table}

If iceberg trade costs were the primary source of imbalances, one might expect our results on the relationship between trade deficits to be even stronger.  The idea behind this view being that tariffs might be necessary to offset non-tariff trade barriers that might be suppressing exports from a country that has a deficit.  Table \ref{tbl:Optimal_Taus_Full_iceberg} reports optimal unilateral tariffs and Nash equilibrium tariffs, under the assumption that bilateral imbalances are due to iceberg trade costs.  For the free trade baseline, we remove tariffs, but we do not eliminate iceberg trade costs.  We find that our results are largely unchanged, suggesting that the cause of trade imbalances is less important for trade policy than the fact that they exist.

\subsection{Alternative Method for Eliminating Aggregate Trade Imbalances}
\label{subsec:no_agg_imbalances}
In static trade models, there are two common strategies to deal with aggregate trade imbalances.  The first, as we did in Section \ref{sec:calibration} is to calibrate the model to match observed aggregate trade imbalances, and include the trade deficit, $D$, as a transfer in the consumer budget constraint, in equation \eqref{BC2}.  The other way to handle aggregate trade imbalances is to focus on scenarios in which there are no aggregate trade imbalances.  In the main paper, we accomplished this by calibrating the model to the data, and then solving for a counterfactual equilibrium with $D_{i}=0$, and using this counterfactual as our baseline.  An alternative approach for eliminating aggregate deficits is to rebalance imports or exports for each country with the rest of the world, creating synthetic trade flows, $\widetilde{p_{{\scriptscriptstyle \text{ROW}},j}^s y_{{\scriptscriptstyle \text{ROW}},j}^s}$, that ensure trade is balanced for each country on aggregate.  We implement this second method by scaling exports to the rest of the world proportionally to eliminate imbalances. Specifically, we construct,

\begin{equation}
    \widetilde{p_{{\scriptscriptstyle \text{ROW}},j}^s y_{{\scriptscriptstyle \text{ROW}},j}^s} = p_{{\scriptscriptstyle \text{ROW}},j}^s y_{{\scriptscriptstyle \text{ROW}},j}^s \left(1 + D_{j}\left(\frac{p_{\scriptscriptstyle \text{ROW},j}^s y_{\scriptscriptstyle \text{ROW},j}^s}{\sum_{s}{p_{{\scriptscriptstyle \text{ROW}},j}^s y_{{\scriptscriptstyle \text{ROW}},j}^s}} \right) \right).
\end{equation}
We then substitute $\widetilde{p_{{\scriptscriptstyle \text{ROW}},j}^s y_{{\scriptscriptstyle \text{ROW}},j}^s}$ in for $p_{{\scriptscriptstyle \text{ROW}},j}^s y_{{\scriptscriptstyle \text{ROW}},j}^s$ when calibrating the model to match the data.  This ensures that $D_{j}=0$ and there are no aggregate imbalances.

\begin{table}[htb]
\caption{Optimal Tariffs and Welfare Changes by Model Structure  \\ (No Aggregate Trade Imbalances)}
\label{tbl:Multisector_Full_noAgg_results}
\begin{center}
\begin{tabular}{cccSS|SSSS}
\toprule
\multicolumn{3}{c}{} & \multicolumn{2}{c}{Unilateral Tariffs} & \multicolumn{4}{c}{Nash Tariffs} \\
\cmidrule(lr){4-5} \cmidrule(lr){6-9} 
& &  & \multicolumn{2}{c}{USA} & \multicolumn{2}{c}{USA} & \multicolumn{2}{c}{CHN} \\
IO & Services & \shortstack{Free\\Trade} & {$\overline{\tau^{*}}$} & {$\Delta W$(\%)} & {$\overline{\tau^{*}}$} & {$\Delta W$(\%)} & {$\overline{\tau^{*}}$} & {$\Delta W$(\%)} \\ 
\cmidrule(lr){1-3} \cmidrule(lr){4-5} \cmidrule(lr){6-7} \cmidrule(lr){8-9}
 No & No & No & 9.95 & 0.050 & 10.04 & 0.044 & 10.71 & -0.059 \\ 
 No & Yes & No & 9.32 & 0.012 & 9.66 & 0.010 & 10.79 & -0.033 \\ 
 Yes & No & No & 10.34 & 0.088 & 10.44 & 0.072 & 11.54 & -0.155 \\ 
 Yes & Yes & No & 9.23 & 0.020 & 9.58 & 0.016 & 11.61 & -0.096 \\ 
\bottomrule
\end{tabular}
\end{center}
\end{table}

After rebalancing to eliminate aggregate trade imbalances, we recompute optimal unilateral and Nash equilibrium tariffs.  These results are shown in Table \ref{tbl:Multisector_Full_noAgg_results}. We find that the results are largely unchanged, suggesting that the presence of aggregate trade imbalances is not crucial for our results, and that bilateral imbalances are quantitatively more important.  This may be due to the fact that, empirically, aggregate imbalances are relatively small as a share of GDP, as shown in Figure \ref{Fig--Global_Imbalances_GDP_Agg}, whereas, comparatively, bilateral imbalances are much larger as a share of total trade, as shown in Figure \ref{Fig--Global_Imbalances}.

\subsection{Numerical Elasticities}
\label{subsec:NumElast}
We compute the numerical elasticity, $\varsigma_{ij}^{s}$ in each sector for both U.S. imports from China and the reverse.  To compute numerical elasticities, we perturb the baseline by adding 0.0001 to the iceberg trade cost, $t_{ij}^s$, in that sector.  We then solve the counterfactual equilibrium, and use the following formula,
\begin{equation}
    \varsigma_{ij}^{s} = \frac{\log{y_{ij}^{s'}}-\log{y_{ij}^{s}}}{\log{p_{ij}^{s'}}-\log{p_{ij}^{s}}},
\end{equation}
where $y_{ij}^{s'}$ is the output in the counterfactual equilibrium following perturbation.

Table \ref{tbl:Num_Elasticity_results} reports the mean numerical elasticity across each of the 22-sectors in our quantitative application from Section \ref{sec:application}. The numerical elasticities are computed at the baseline and reported for eight different scenarios (whether bilateral or deficits are present and whether the benchmark has free-trade or observed tariffs).  We also report the mean absolute and percentage difference between the numerical elasticities for the United States and China.  For the percentage difference, we use the Armington elasticity as the denominator.  The negative numbers indicate that US imports from China are more inelastic compared to Chinese imports from the United States.   These results show that the intuition from the illustrative model from Section \ref{sec:toy_model} applies to our fully specified framework.  In our calibrated model, bilateral and aggregate trade imbalances make it so that U.S. import demand is relatively less elastic compared to Chinese import demand.  This explains why the United States is able to experience higher welfare gains (or lower welfare losses) from a trade war.

\begin{table}[htb]
\caption{Optimal Tariffs and Welfare Changes (With and Without Deficits)}
\label{tbl:Num_Elasticity_results}
\begin{center}
\begin{tabular}{cccSSSS}
\toprule
\multicolumn{3}{c}{}& \multicolumn{2}{c}{Mean $\varsigma_{ij}^{s}$} & \multicolumn{2}{c}{Mean Difference} \\
\cmidrule(lr){4-5} \cmidrule(lr){6-7} 
\shortstack{Bilateral\\Deficit}  &  \shortstack{Aggregate\\Deficits}  & \shortstack{Free\\Trade} & {USA} & {CHN}  & {Absolute} & {Percentage} \\
\cmidrule(lr){1-3} \cmidrule(lr){4-5} \cmidrule(lr){6-7} 
 Yes & Yes & No & 5.47 & 5.71 & -0.24 & -3.87 \\ 
 Yes & No & No & 5.52 & 5.70 & -0.18 & -2.97 \\ 
 No & Yes & No & 5.48 & 5.53 & -0.05 & -1.29 \\ 
 No & No & No & 5.51 & 5.57 & -0.06 & -1.35 \\ 
\cmidrule(lr){1-3} \cmidrule(lr){4-5} \cmidrule(lr){6-7} 
 No & No & Yes & 5.43 & 5.69 & -0.26 & -4.22 \\ 
 Yes & No & Yes & 5.48 & 5.68 & -0.20 & -3.27 \\ 
 No & Yes & Yes & 5.44 & 5.52 & -0.08 & -1.72 \\ 
 No & No & Yes & 5.47 & 5.56 & -0.09 & -1.75 \\ 
\bottomrule
\end{tabular}
\end{center}
\end{table}

\subsection{Aggregation of Sectors}
\label{subsec:aggregation}
In our fully-specified model with 22 tradable goods-producing sectors, it is implausible to compute Nash equilibria using traditional approaches that rely on an exhaustive search over possible strategies.  Discretizing the set of possible tariff rates to $N$ points means that we would need to evaluate $22^N$ possible combinations for each country to compute its best response.  Even for an small grid with $N=3$ points per industry, this would require computing over 31 billion combinations, each of which requires solving for a counterfactual equilibrium in order to determine Welfare payoffs.  For models with a small number of sectors, however, it can be possible to exhaustively search over possibilities to establish a Nash equilibrium and show that our genetic algorithm converges to the correct solution.  We do this for a single sector model, and verify that our genetic algorithm indeed converges to the same solution as the exhaustive search.  In Section \ref{subsec:uniform_tariffs} we present a graph of the welfare consequences of deviations from the Nash Equilibrium uniform tariffs; the graph for one sector model looks similar.

Aggregation is straightforward, and involves combining sectors into a smaller number of sectors by adding together their trade flow values and input-output into the newly aggregated composite sectors.  For tariff rates in the baseline economy, we average the reported tariff rates across each of the original sectors to arrive at a tariff rate for the composite sector.  Aggregation can also be implemented across countries and regions; for example, by combining individual European Union countries into an EU region aggregate --- a step we take in our main analysis to accelerate the computation of each counterfactual equilibrium.  After aggregating the data, we simply calibrate our model to this aggregated data, and no further changes are needed.  %

\begin{table}[htb]
\caption{Optimal Tariffs and Welfare Changes by Model Structure \\ (Single Aggregated Tradable Goods Producing Sector)}
\label{tbl:Small_results}
\begin{center}
\begin{tabular}{cccSS|SSSS}
\toprule
& & & \multicolumn{2}{c}{Unilateral Tariffs} & \multicolumn{4}{c}{Nash Tariffs} \\
\cmidrule(lr){4-5} \cmidrule(lr){6-9} 
& &  & \multicolumn{2}{c}{USA} & \multicolumn{2}{c}{USA} & \multicolumn{2}{c}{CHN} \\
IO & Services & \shortstack{Free\\Trade} & {$\overline{\tau^{*}}$} & {$\Delta W$(\%)} & {$\overline{\tau^{*}}$} & {$\Delta W$(\%)} & {$\overline{\tau^{*}}$} & {$\Delta W$(\%)} \\ 
\cmidrule(lr){1-3} \cmidrule(lr){4-5} \cmidrule(lr){6-7} \cmidrule(lr){8-9}
 No & No & No & 14.25 & 0.113 & 14.43 & 0.054 & 17.90 & -0.137 \\ 
 No & Yes & No & 12.96 & 0.025 & 13.09 & 0.010 & 18.09 & -0.075 \\ 
 Yes & No & No & 13.62 & 0.167 & 13.83 & 0.053 & 19.00 & -0.335 \\ 
 Yes & Yes & No & 11.01 & 0.030 & 11.11 & 0.007 & 17.77 & -0.178 \\ 
\cmidrule(lr){1-3} \cmidrule(lr){4-5} \cmidrule(lr){6-7} \cmidrule(lr){8-9}
 No & No & Yes & 12.85 & 0.161 & 13.11 & 0.053 & 13.85 & -0.131 \\ 
 No & Yes & Yes & 10.96 & 0.032 & 11.12 & 0.008 & 13.73 & -0.065 \\ 
 Yes & No & Yes & 12.09 & 0.236 & 12.35 & 0.040 & 14.60 & -0.312 \\ 
 Yes & Yes & Yes & 10.05 & 0.049 & 10.19 & 0.005 & 13.87 & -0.175 \\ 
\bottomrule
\end{tabular}
\end{center}
\end{table}

Table \ref{tbl:Small_results} reports optimal tariff rates and corresponding welfare changes for model with a single tradable goods-producing sector and, when applicable, a single service sector.  We set the Armington elasticity in each of these composite sectors to 5.5.  The optimal and Nash equilibrium tariff rates differ slightly from those in Table \ref{tbl:Multisector_Full_results}, where our model has 22 tradable goods-producing sectors; however, the qualitative results and key insights remain similar.  In both, the United States expects small welfare gains from optimal unilateral or Nash equilibrium tariffs relative to the pre-trade war baseline. Furthermore, accounting for service sectors and input-output linkages remains crucial for the sign of the United States' welfare changes relative to a free trade baseline.  %

\section{Appendix: Description of Genetic Algorithm}
\counterwithin{table}{section}
\counterwithin{figure}{section}
\counterwithin{algorithm}{section}
\setcounter{algorithm}{0}

\label{subsec:ga}
Genetic algorithms are a type of global optimization heuristic that fall within the broader class of evolutionary algorithms.  Evolutionary algorithms are so named because they maintain a population of solution candidates, which they update based on principles inspired by survival of the fittest.  Genetic algorithms work by generating the next iteration (generation) of solution candidates (children) through the combination (crossover) and alteration (mutation) of sub-components (genes) of previous solution candidates (parents), where solution candidates with high fitness are more likely to serve as parents to future iterations compared to low fitness candidates (selection).  The terms in parenthesis are the official terms used within the genetic algorithm literature and highlight the biological motivation underlying the algorithm.  Although the components of selection, crossover, and mutation are core to the concept of genetic algorithms, there is a wide range of methods for implementing each of these features.  We refer readers to \citet{mitchellIntroductionGeneticAlgorithms1998} for a broad overview of genetic algorithms and \citet{Crepinsek2013} for a discussion of how these choices can be framed in terms of trade-offs between exploration of the solution space to avoid local optima and exploitation of high fitness candidates to speed solution refinement and convergence.  Since researchers and applications differ in terms of how these trade-offs are valued, there is no universal best approach, although the algorithm can be fine tuned to improve performance for specific problems.  Below we discuss our specific implementation of the genetic algorithm.

Our genetic algorithm is outlined in Algorithm \ref{alg:genetic}.  In both the algorithm and this discussion, we set country indices to correspond to the problem of the United States setting tariffs on imports from China to maximize U.S. Welfare, taking as given China's tariffs set on imports from the United States.  We do this to aid in readability.  The algorithm follows identically after updating countries indices for the reverse problem (optimal tariffs for China) and for applications involving trade wars between two arbitrary countries.

\begin{algorithm}
    \caption{Genetic Algorithm for Computing Best Response Tariff Rates}
    \label{alg:genetic}
    \begin{algorithmic}[1]
    \renewcommand{\algorithmicrequire}{\textbf{Setup:}}
    \renewcommand{\algorithmicensure}{\textbf{Algorithm:}}
    \REQUIRE {\it Country indices are set for USA's optimal tariff rates for improved readability.}
        \STATE Calibrate model to match baseline economy according to Section \ref{subsec:calibration}
\\ \textbf{•} \textit{Note: Chinese tariffs on imports from China, $\tau_{CHN,USA}^{s}$, taken as given and affect solution}\\
         \STATE Set bounds for tariffs \( \boldsymbol{\tau}_{USA} \in [\underline{\tau},\overline{\tau}] = [1.00,5.00] \)  (\textit{i.e., max rate of 400 percent}) \\
	\STATE Discretize solution space to four decimal places (\textit{hundredths of a percentage point})  \\
	\STATE Initialize a population $P=160$ of solution candidates for generation $g=0$
\\ \textbf{•} The first $P/2$  candidates are selected from a random uniform distribution \(  U [\underline{\tau},\overline{\tau}] \)
\\ \textbf{•}  \textit{Optional: Replace any number of these draws with user-specified candidates}  
\\ \textbf{•} The remaining $P/2$ candidates are the opposites of these draws 
    \ENSURE  \textit{Estimation of Best Response Tariff Rates}
    \STATE Evaluate the fitness of each candidate, $W_{USA}^{\{p,0\}}$. Set $g=G_{total}=0$ and $G_{stall}=0$.
    \WHILE{ $G_{stall} \le \textit{Stall Limit} (=50)$ \&  $G_{total} \le \textit{Total Limit} (=1000)$   } %
\PROCEDURE[Fitness Scaling]{(Rank Fitness Scaling)}{Compute candidate fitness: $F^{\{p,g\}}$}
	\STATE  Rank candidates $\boldsymbol{\tau}_{USA}^{\{p,g\}}$ by $W_{USA}^{\{p,g\}}$ (rank 1 for highest welfare).
	\STATE  Compute $F^{\{p,g\}}\equiv 1/\sqrt{r^{\{p,g\}}}$, where $r^{\{p,g\}}$ is rank of $p$ within $g-1$
	\STATE Include top $P_{elite}=16$ ranked candidates as elite children
	\ENDPROCEDURE
\STATE \textsc{Roulette Selection:} Parents are randomly selected for crossover and mutation with probability in proportion to $F^{\{p,g\}}$
 \PROCEDURE[Crossover]{(Laplace Crossover)}{Create $P_{X}=116$ children from crossover}
	\STATE Randomly select two parents, $p_1$ and $p_2$, for each child
	\STATE Create child $c$ from its parents according to Laplace crossover \eqref{eq:laplace_crossover}
	\ENDPROCEDURE	
\PROCEDURE[Mutation]{(Adaptive Power Mutation)}{Create $P_{mut}=28$ children from mutation}
        \STATE Select one parent, $p$, for each child
        \STATE Create child $c$ according to adaptive power mutation according to \eqref{eq:adaptive_power_mutation:step_size}--\eqref{eq:adaptive_power_mutation:final}
	\ENDPROCEDURE	
    \STATE Increment $g$ by one and set ${\{p,g\}} = {\{c,g-1\}}$.
    \STATE Evaluate the fitness of each candidate, $W_{USA}^{\{p,g\}}$
    \STATE Compute the max welfare $W_{USA}^{\{*,g\}}=\max_{p} {W_{USA}^{\{p,g\}}}$ over candidates in generation $g$
 \STATE Compute the geometric weighted mean change, $ \nabla^{\{g\}}$, according to \eqref{eq:geometric_weighted_change} %
\IF{$\nabla^{\{g\}}\le \textit{tolerance}$ (where $\textit{tolerance}=10^{-4})$}
\STATE Increment $G_{stall}$ by 1
\ELSE
\STATE Reset $G_{stall}=0$
\ENDIF
\STATE  Increment $G_{total}$ by 1
    \ENDWHILE
    \RETURN $\boldsymbol{\tau}_{USA}^{*} \equiv \boldsymbol{\tau}_{USA}^{\{*,G\}}$ and $W_{USA}^{*} \equiv W_{USA}^{\{*,G\}}$
    \end{algorithmic} 
\end{algorithm}

Pre-initialization of the algorithm involves calibrating the model to match a baseline economy, as specified in Section \ref{sec:calibration}.  We then initialize a population, $P$, of solution candidates, \(  \boldsymbol{\tau}_{USA}^{\{p,g\}}  \equiv \{ \tau_{USA,CHN}^1,...,\tau_{USA,CHN}^{22}\}^{\{p,g\}} \), where $ p\in P $ represents candidate $p$ of generation $g$.  After including any user-specified seed candidates --- for example from a previous execution of the genetic algorithm --- we randomly generate candidate solutions until we have $P/2$ candidates (including seeds).  We generate these candidates by taking independent draws for each component, $\tau_{USA,CHN}^{ss}$, from a uniform distribution over the interval $[\underline{\tau},\overline{\tau}] = [1.0,5.0]$, where $\underline{\tau}$ and $\overline{\tau}$ are the lower and upper bounds on the tariff rates, respectively.  The remaining $P/2$ candidates are the opposites of these draws, where the opposite of  \(  \boldsymbol{\tau}_{USA}^{\{p,g\}} \) is  \( \left( \overline{\tau} + \underline{\tau} - \boldsymbol{\tau}_{USA}^{\{p,g\}} \right) \).  This ensures that the initial population is spaced out across the potential solution space and aids in initial exploration.  Restricting the solution space by using bounds on tariff rates both speeds convergence and helps to avoid the possibility of multiple equilibria which could exist for very high tariff rates as discussed in Section \ref{sec:optimal_tariffs_algorithm}.  Setting $\underline{\tau}=1.0$ rules out subsidies as a policy measure, although it would be straightforward to relax this restriction.  We discretize the solution space to four decimal places, which corresponds to computing optimal tariff rates to hundredths of a percentage point, and further speeds convergence.

After generating an initial population of solution candidates, we evaluate welfare for each candidate, $W_{USA}^{\{p,g\}}$. Welfare is computed by solving the model for the counterfactual equilibrium at the candidate tariff rates, and then computing the welfare of the United States in this equilibrium according to the expression in \eqref{eq:optimal_tariffs}.  After computing welfare, we rank each candidate within a generation according to its welfare, where rank 1 is assigned to the candidate with the highest welfare.  We randomly select candidates from the current generation to serve as parents for creating children candidates for the next generation, which occurs through crossover and mutation.

We create $P_{X}=116$ children from crossover using Laplace crossover \citep{Chuang2015}.  Laplace crossover involves selecting two parents for each child, $p_1$ and $p_2$.  We implement the selection of parents according to inverse rank fitness scaling and roulette selection.  Inverse rank fitness scaling involves computing $F^{\{p,g\}}\equiv 1/\sqrt{r^{\{p,g\}}}$, where $r^{\{p,g\}}$ is the rank of candidate $p$ in generation $g$, and is used to encourage exploration by ensuring selection is not restricted to only the highest fitness candidates.  Roulette selection means that each candidate will be selected with probability in proportion to its scaled fitness score.  For a given child with multiple parents, each parent is selected without replacement.  Across children, however, parents are selected with replacement.   We use this same selection procedure for selecting parent candidates for mutation, which we discuss later in this section.

For a given child, $c$, Laplace crossover involves the componentwise combination of its parents, $p_1$ and $p_2$, according to
\begin{equation}
    \begin{aligned}
{\tau_{USA,CHN}^{s}}^{\{c,g\}}&={\tau_{USA,CHN}^{s}}^{\{p_{1},g\}} \\
&+\lambda_{s}^{\{c,g\}}\left| {\tau_{USA,CHN}^{s}}^{\{p_{1},g\}}-{\tau_{USA,CHN}^{s}}^{\{p_{2},g\}}\right|,
\end{aligned}
\label{eq:laplace_crossover}
\end{equation}
where $\lambda_{s}^{\{c,g\}}$ is drawn from a Laplace distribution with density function
\begin{equation*}
f(x|a,b)=\frac{1}{2b} e^{-{\left|x-a\right|}/b}.
\end{equation*}
We set $a=0$ to center the child's components around the corresponding component for $p_1$, and $b$ adjusts the spread of the distribution.  We set $b=0.35$, which implies around 75 percent of the children's components will be closer to $p_1$ than to $p_2$.  Random draws of the Laplace distribution are generated by computing $\lambda_{s}^{\{c,g\}}=a- \text{sign}(u) b \log{\left(1-2\left|u\right|\right)}$ where $u$ is from a standard uniform distribution and $\text{sign}()$ is the signum function.  To enforce children to conform to our constrained solution space, $ [\underline{\tau},\overline{\tau}] = [1.00,5.00]$, we re-generate $\lambda_{s}^{\{c,g\}}$ for sectors that would exceed these bounds.

We also create $P_{mut}=28$ children from mutation using adaptive power mutation.  Power mutation involves selecting a single parent, $p$, for each child, $c$, and then applying a mutation based on increasing or decreasing each component by an amount equal to a uniform draw raised to a power \citep{Deep2007}, scaled by each component's distance to the boundary.  Adaptive power mutation indicates the strength of mutation --- the power --- varies depending on whether a candidate has fitness greater than or below average fitness.  The step size for each component is given by
\begin{equation}
    \delta_{mut}^{\{c,g\}}=\mathbb{1}_{fit}^{{\{p,g\}}}u_{step}^{10} + \left(\sim\mathbb{1}_{fit}^{{\{p,g\}}}\right)u_{step}^{4},
    \label{eq:adaptive_power_mutation:step_size}
    \end{equation}
where $u_{step}$ is drawn from an independent standard uniform distribution for each component, and $\mathbb{1}_{fit}^{{\{p,g\}}}$ is an indicator function that is equal to 1 if the fitness of candidate $p$ in generation $g$ is above average fitness (i.e., with rank greater than $P/2$), and equal to 0 otherwise.  Note this indicator does not vary by component.  When the indicator function is one, we apply a power mutation with power equal to 10, which means approximately 37 percent of step sizes will be greater than 1 percentage of the distance to the boundary.  When $\mathbb{1}_{fit}^{{\{p,g\}}}$ is zero, the logical complement, $\sim\mathbb{1}_{fit}^{{\{p,g\}}}$ will be equal to zero, and we apply a power mutation with power 4, which means approximately 68 percent of step sizes will be greater than 1 percentage of the distance to the boundary.  An adaptive approach allows for improved exploitation for good solutions while maintaining the ability to make larger changes for lower fitness candidates.

The direction of mutation (increase or decrease) is likewise determined probabilistically, based on the scaled distance of each parameter from constraint bounds. Parameters closer to their bounds are more likely to move away from them, up to a maximum rate of 0.75 of the time.  For this, we define a componentwise indicator function $\mathbb{1}_{dir}^{{\{p,g\}}}$ that is equal to 1 if a random standard uniform draw, $u_{dir}$, is less than the scaled distance of the component from the constraint bounds, i.e.,
\begin{equation}
    \mathbb{1}_{dir}^{{\{p,g\}}} = 1 \text{ if } u_{dir}<\min\left(\max{\left(\frac{{\tau_{USA,CHN}^{s}}^{\{p,g\}}-\underline{\tau}}{\overline{\tau}-\underline{\tau}},0.25\right)},0.75\right),
\end{equation}
and equal to zero otherwise.  The complete mutation is applied by adjusting parent component values using the step size and direction indicator function according to
\begin{equation}
    \begin{aligned}
    {\tau_{USA,CHN}^{s}}^{\{c,g\}}={\tau_{USA,CHN}^{s}}^{\{p,g\}}&-\delta_{mut}^{\{c,g\}} \mathbb{1}_{dir}^{{\{p,g\}}} \left({\tau_{USA,CHN}^{s}}^{\{p,g\}}-\underline{\tau}\right) \\
    &+\delta_{mut}^{\{c,g\}}\left(\sim\mathbb{1}_{dir}^{{\{p,g\}}}\right)\left(\overline{\tau}-{\tau_{USA,CHN}^{s}}^{\{p,g\}}    \right).
    \end{aligned}
    \label{eq:adaptive_power_mutation:final}
\end{equation}

After both crossover and mutation, we round children candidates to four decimal places.  We increment $g$ by one and set ${\{p,g\}} = {\{c,g-1\}}$.  Across each generation, we compute the welfare of the top performing candidate $W_{USA}^{\{*,g\}}=\max_{p} {W_{USA}^{\{p,g\}}}$.  We evaluate the step-weighted geometric mean  increase in  $W_{USA}^{\{*,g\}}$ across generations, where the weight for each generation's increase in welfare is proportional to $(1/2)^{t}$ where $t$ is how many generations in the past a given increase.  We denote the step-weighted geometric mean increase by  %
\begin{equation}
 \nabla^{\{g\}} \equiv  {\left( \prod_{t=0}^{g-1} {\left| W_{USA}^{*\{g-t\}}-W_{USA}^{*\{g-(t+1)\}} \right|}^{\left( 1/2 \right)^{t}} \right)}^{1 / {\sum_{t=0}^{g-1}{{\left( 1/2 \right)}^t}}}.
\label{eq:geometric_weighted_change}
\end{equation}
Each generation automatically includes the $P_{elite}$ candidate solutions, called elite children, with the highest corresponding welfares from the previous generation.  This implies that $ W_{USA}^{*\{g\}}$ is monotonically increasing with $g$.  Note also that we do not need to recompute counterfactual welfare for elite children in future generations, as these candidates and their welfare will be unchanged. 

Each time the step-weighted geometric mean increase in welfare, $\nabla^{\{g\}}$, falls below our pre-specified tolerance level, $10^{-4}$, we increment a stall counter, $G_{stall}$, that tracks how many generations have passed since the last significant improvement in welfare.  If $\nabla^{\{g\}}$ rises above the tolerance level, we reset the stall counter by setting $G_{stall}$ equal to zero.   We also increment a counter, $G_{total}$, that tracks the total number of generations that have passed.  We continue to iterate the genetic algorithm until $G_{stall}$ exceeds our pre-specified limit of 50, or until $G_{total}$ exceeds our pre-specified limit of 1000.  The algorithm terminates with the best performing solution candidate, $\boldsymbol{\tau}_{USA}^{*}$, and its corresponding welfare, $W_{USA}^{*}$; which we take as our solution to \eqref{eq:optimal_tariffs}.

We further accelerate convergence of the genetic algorithm by using a hybrid approach that combines the genetic algorithm with a local optimization subroutine.  To implement a hybrid genetic algorithm, we follow the same steps as in the genetic algorithm described by Algorithm \ref{alg:genetic}, however, we lower the stall limit to 10 and the total limit to 200.  We then apply a local optimization subroutine to each of the top 16 elite candidates from the final generation of the genetic algorithm.  We then re-initialize the genetic algorithm, seeding it with the same top 16 elites and with each unique solution of the local optimization subroutine that yielded improved welfare.  We continue this process until we reach convergence as determined by \( \max{ (\lVert \Delta  \boldsymbol{\tau}_{i}^{*} \rVert,\lVert\Delta W_{i}^{*}\rVert) } \geq \textit{tol} \), which more closely matches the termination criteria of Algorithm \ref{alg:br_tau}.  This hybrid approach allows us to exploit the genetic algorithm's ability to explore the solution space and avoid local optima, while also exploiting the local optimization subroutine's ability to refine solutions.  Even without the hybrid approach, choosing smaller stall limits and re-initializing the algorithm multiple times can yield improved performance.  As before, we do not emphasize the importance of any particular local optimization heuristic.  We implement an interior trust region local optimization algorithm with bounds based on \citet{Coleman1996}, and find it speeds convergence and accuracy.  Attempting to bypass the genetic algorithm altogether, however, and relying solely on these local optimization heuristics can be expected to lead to poor performance for models with more than a small number of sectors.%

\renewcommand{\thesection}{\Alph{section}}
\renewcommand{\thesubsection}{\Alph{subsection}}

\end{document}